\documentclass[12pt]{article}
\usepackage[top=1in,bottom=1in,left=1in,right=1in]{geometry}
\pdfoutput = 1
\usepackage[noEucal]{main}
\usepackage{aas_macros}
\usepackage[utf8]{inputenc}
\makeatletter \g@addto@macro\@floatboxreset\centering \makeatother
\usepackage{graphicx}
\usepackage[scr=boondoxo]{mathalfa}
\usepackage[bbgreekl]{mathbbol}
\DeclareSymbolFontAlphabet{\mathbbm}{bbold}
\usepackage[font={small,it}]{caption}

\def\Im{{\text{Im}\,}}

\newcommand{\ted}{\text{d}}
\newcommand{\bift}[1]{\big( #1\big)}

\newcommand{\Bitd}[1]{\Big[ #1\Big]}
\newcommand{\Bift}[1]{\Big( #1\Big)}
\begin{document}
\begin{titlepage}
\unitlength = 1mm

\vskip 3cm
\begin{center}
\openup .5em

{\Large{Density of states of quantum systems from free probability theory: a brief overview}}

\vspace{0.8cm}

Keun-Young Kim$^\ddagger$$^{\Diamond}$\footnote{fortoe@gist.ac.kr},  Kuntal Pal$^\Diamond$\footnote{kuntalpal@gist.ac.kr} 

\vspace{1cm}
{\it $^\Diamond$ Department of Physics and Photon Science, Gwangju Institute of Science and Technology, \\ 123 Cheomdan-gwagiro, Gwangju 61005, Korea}\\
{\it $^\ddagger$ Research Centre for Photon Science Technology, Gwangju Institute of Science and Technology,\\ 123 Cheomdan-gwagiro, Gwangju 61005, Korea}

\vspace{0.8cm}
\begin{abstract}

We provide a brief overview of approaches for calculating the density of states of quantum systems and random matrix Hamiltonians using the tools of free probability theory. For a given Hamiltonian of a quantum system or a generic random matrix Hamiltonian, which can be written as a sum of two non-commutating operators, one can obtain an expression for the density of states of the Hamiltonian from the known density of states of the two component operators by assuming that these operators are mutually free and by using the free additive convolution. In many examples of interacting quantum systems and random matrix models, this procedure is known to provide a reasonably accurate approximation to the exact numerical density of states. We review some of the examples that are known in the literature where this procedure works very well, and also discuss some of the limitations of this method in situations where the free probability approximation fails to provide a sufficiently accurate description of the exact density of states. Subsequently, we describe a perturbation scheme that can be developed from the subordination formulas for the Cauchy transform of the density of states and use it to obtain approximate analytical expressions for the density of states in various models, such as the Rosenzweig-Porter random matrix ensemble and the Anderson model with on-site disorder.  

\end{abstract}

\vspace{1.0cm}
\end{center}
\end{titlepage}
\pagestyle{empty}
\pagestyle{plain}

\tableofcontents

\section{Introduction}

The purpose of this manuscript is to collect and expand some of the results on the density of states (DOS) or the density of eigenvalues of realistic quantum mechanical systems and random matrices using the free probability theory that have been obtained so far by various authors in different works, with the goal of introducing them in a, hopefully, coherent manner to the readers who are unfamiliar with, but interested in the subject. Since the DOS of the Hamiltonian determines the equilibrium properties of the quantum system under consideration through its relation with the partition function, it is one of the primary quantities of interest that plays an important role in probing the physics of quantum systems, quantifying quantum information and correlations, and understanding quantum signatures of chaos, among others. 
In these notes, our goal is to review the use of a specific mathematical theory of highly non-commuting operators, namely, the free probability theory, to gain insights into this important physical problem of describing the DOS of complex quantum systems and generic random matrix Hamiltonians.

The free probability theory  \cite{voiculescu1992free, speicher2019lecture, nica2006lectures} is a branch of mathematics that deals with non-commutative random variables in a way that is analogous to the concept of independence of random variables in classical probability theory, by introducing the notion of freeness or free independence. Since its introduction in the context of operator algebras by D. Voiculescu, free probability theory has emerged as an indispensable tool in several areas of mathematics and physics, specifically in the theory of random matrices \cite{voiculescu1991limit, mingo2017free}. 
For a selection of applications of free probability theory in the physics literature, we refer to: 1. \cite{Gopakumar:1994iq, Douglas:1994zu, zee1996law, janik1997various, Engelhardt:1996da} for applications in large $N$ gauge theories and random matrix models; 2. \cite{Pappalardi:2022aaz, Pappalardi:2023nsj, PhysRevX.15.011031, Vallini:2024bwp, Fullgraf:2025tsw} for the use of free probability in understanding the generalised eigenstate thermalisation hypothesis and the applications of the free cumulants in that context; 3. \cite{Hruza} for the use of free cumulants in analysing the dynamics of fluctuations in mesoscopic systems; 4. In non-hermitian random matrix theories \cite{janiknonh, burda}; 5. \cite{Chen:2024free, Jindal:2024zcg, Camargo:2025zxr} for connections with quantum chaos and quantum ergodic theory, and the emergence of free probability prediction of eigenvalue density of a sum of two operators, one of which is time evolved by chaotic quantum mechanical Hamiltonians \cite{Camargo:2025zxr}.\footnote{This list of references is quite incomplete. We have mentioned the works we are familiar with; besides, these are only references from physics literature. For various other works, specifically, on random unitary circuits, free Lévy matrices, and understanding driven topological phases, we refer to \cite{Fritzsch:2025arx, Dowling:2025cxr}, \cite{burdalevy}, and   \cite{Shtanko:2018oih}, respectively. See \cite{collins2016random, Wang:2022ots, Vardhan:2025rky } for quantum information-theoretic applications of free probability, and \cite{Wu:2023kin} for its use to gain insights into the double scaling limit of the Sachdev–Ye–Kitaev model (see also \cite{pluma2022dynamical}).}   

As mentioned above, in these notes, our aim is to review the application of free probabilistic techniques to a particular physical problem, namely, the computation of the DOS of quantum Hamiltonians with large dimensions.
In free probability theory, there exists a procedure that is the counterpart of the usual additive convolution of the probability distribution functions in classical probability, known as the free additive convolution, which can be utilised to obtain the DOS of a sum of two non-commutating operators.  
Specifically, we consider the Hamiltonian of a quantum system, possibly with disorder and interaction terms, of the form $\BA+\BB$, where $\BA$ and $\BB$ are two generic non-commuting operators, whose DOS are assumed to be given. In a generic scenario, since the operators $\BA$ and $\BB$ do not share common eigenvectors, the DOS of the Hamiltonian is difficult to obtain, even if the individual DOS are known. To this end, a useful strategy one can adopt to make progress is to assume that the `complicated' operators $\BA$ and $\BB$ are such that their eigenvectors are randomly rotated with respect to one another - a situation which is canonical in free probability theory, and specific results proven in mathematics literature imply that, the operators $\BA$ and $\BB$ are asymptotically free in the limit when their dimensions are sufficiently large. Then one can employ the full machinery of the free additive convolution (the counterpart of classical additive convolution in free probability theory) to study the DOS of the Hamiltonian. This strategy has been adopted in several works to study the DOS of complex quantum systems and random matrix Hamiltonians (see, e.g.,  \cite{neu1995random, Chenprl, movassagh2010isotropic, gudowska2003free, gudowska1998bridged, Pollock:2025acf, Venturelli:2022hka, Jahnke:2025exd}), and it has been shown that free probability approximation can provide a very good description of the exact numerical DOS. 

Since, the mathematical concept of freeness requires the matrices under considerations to be infinite dimensional (in the related concept of asymptotic freeness, one needs to consider the limit when the dimension of a sequence of matrices $N \rightarrow \infty$), whereas the operators that arise in the context of realistic quantum systems are finite dimensional, one can not expect the quantum mechanical operators $\BA$ and $\BB$ to be free in the strict mathematical sense. As we shall discuss, even though one can construct models where these operators are asymptotically free by construction, in realistic physical systems, the freeness is expected to be valid up to a certain `order'.
A natural question one can ask in this context is: for which systems does the free probability approximation work, and how good is the free probability approximation of the DOS of realistic quantum systems with disorder and interactions? We shall review several cases where free probability approximation provides a very good description of the numerical DOS, as well as discuss situations in which it fails to provide a satisfactory match with the exact numerical DOS. 

We start in the Section \ref{sec_freeness} by recalling the basic definitions of freeness and asymptotic freeness, and describing the procedure of free additive convolution of the  DOS of two non-commutating operators by introducing the $R$-transform \cite{voiculescu1986addition, voiculescu1992free, speicher2019lecture} and the associated subordination formulas for the Cauchy transform. In section \ref{sec_dos_quantum}, we review several cases where the free convolution is used to obtain the DOS, specifically focusing on the case of a class of local random matrix Hamiltonians introduced in \cite{Pollock:2025acf} and the Anderson model and its various generalisations. We discuss the exactness of the free probability approximation for the DOS in these cases, and also point out situations where it fails to provide an accurate description.

Next, in section \ref{dos_perturb}, we develop a perturbation scheme for calculating the Cauchy transform of the DOS starting from the subordination formulas, and subsequently apply it to the  Rosenzweig-Porter random matrix model, as well as the Anderson model with on-site diagonal disorder. To the best of our knowledge, the results presented in this section for the computation of the DOS  of the Anderson model from the perturbation scheme with different levels of the strength of the interaction have not been previously reported in the literature, and our findings provide an analytical understanding of the DOS of these models. Furthermore, we find a close relationship between the lowest order (in the perturbation parameter) corrections to the Cauchy transform of an initial operator, when the perturbation operator has DOS either a semicircle or an arcsine distribution. 

Next, we present a general discussion of the proposal 
of computing the DOS based on matching the individual moments \cite{movassagh2010isotropic, movassagh2017eigenvalue}, where one can form a convex one-parameter linear combination of fourth moments of classical and free cases (fourth moment is where the moments start to differ)  in section \ref{sec_sum_AB}, and elaborate on an example where a random diagonal matrix is added with a Gaussian random matrix. In section \ref{free_compression}, we review the concepts of free compression of a matrix, and its inverse procedure, known as the free decompression \cite{ameli2025spectral}, and discuss its usefulness in approximately estimating the DOS of a large matrix. To this end, we derive an expression for the Cauchy transform of the DOS of the (de)compressed matrix, and also provide a derivation of the partial differential equation for the same, originally obtained in \cite{ameli2025spectral}. To illustrate this procedure, we explicitly analyse a few examples where the Cauchy transform under free (de)compression can be calculated analytically, and discuss the case where the distribution to be compressed is a free additive convolution of two different distributions. We conclude by providing a summary of the main concepts and the results reviewed/obtained in this paper in section \ref{sec_discussion}.


\section{Freeness and free additive convolution}\label{sec_freeness}

\textbf{Definition of freeness.} Consider the random variables $x_1, x_2, \cdots$, and the polynomials of one variable $p_1(x), p_2(x), \cdots p_m(x)$  (with $m \in \mathbb{N}$), such that $\avg{p_k(x_{i(k)})}=0$ for all $k=1,2, \cdots m$.\footnote{We assume that the random variables under consideration are elements of a $C^*$ probability space, so that it contains a $C^*$ algebra $\CA$ with bounded elements and a linear functional $\varphi: \CA \rightarrow \mrr$ satisfying $\varphi(\mii)=1$ and $\varphi(x^* x)\geq 0$ for all $x \in \CA$. We shall also assume this map to be tracial.} Then the random variables are said to be \textit{freely independent} or free if for all $m \in \mathbb{N}$ \cite{voiculescu1992free, nica2006lectures, mingo2017free},
\begin{equation}\label{free_def}
    \avg{p_1(x_{i(1)})p_2(x_{i(2)})\cdots p_m(x_{i(m)})}=0~,
\end{equation}
for alternating combinations of the random variables, i.e., $i(k) \neq i(k+1)$ for all $k=1,2, \cdots m-1$. As the simplest example, consider two random variables, say, $x, y $ and their polynomials $p_i(x)$ and $q_i(y)$, $i=1,2, \cdots m$ such that $\avg{p_i(x)}=0$ and $\avg{q_i(y)}=0$ $\forall i$. Then $x$ and $y$ are said to be mutually free if for all $m \geq 1$, 
\begin{equation}\label{free_def2}
    \avg{p_1(x)q_1(y)p_2(x)q_2(y)\cdots p_m(x_{})q_m(y)}=0~.
\end{equation}
In both the above definitions $\avg{.}$ denotes a functional ($\varphi(.)$) which maps an element of the $C^*$-algebra to a complex number. For our purposes below, this map can be thought of as the expectation value taken with respect to a suitable quantum state, namely the infinite-temperature thermal state \cite{Camargo:2025zxr}.  From a practical point of view, even though the definition of freeness in \eqref{free_def} and the standard independence of classical commuting random variables are rules for calculating mixed moments of several random variables from the knowledge of individual ones, the concept of freeness is quite different from classical independence.\footnote{Some authors also call the freeness the free independence, since it is the counterpart of the classical independence for non-commutative random variables. Here, we will use the terms "freeness" for non-commutative variables and "independence" for classical (commutative) variables for convenience.} This can be seen easily by considering two random variables, say $x,y$, and computing the mixed fourth moment  $\avg{xyxy}$, which, if the variables are independent 
is simply given by the factorisation $\avg{xyxy}= \avg{x^2}\avg{y^2}$, whereas, if the variables are free, from \eqref{free_def} one can calculate that 
\begin{equation}
    \avg{xyxy} = \avg{x^2}\avg{y}^2+\avg{x}^2\avg{y^2}- \avg{x}^2\avg{y}^2~,
\end{equation}
which, as can be seen, is quite different from the factorisation rule according to independence.\footnote{The first difference between the freeness and the independence arises in the fourth moment, since the first three moments are the same in both situations. } 

Another related concept, which is perhaps more useful for the purpose of this paper, is that of the \textit{asymptotic freeness}, defined as follows. Consider the sequence of random variables $x_{N,1}, x_{N,2}, \cdots$, and the polynomials $p_1(x), p_2(x), \cdots p_m(x)$ for all $m \in \mathbb{N}$, such that $\avg{p_k(x_{N,i(k)})}=0$ for all $k=1,2, \cdots m$. Then the random variables are said to  be asymptotically free if 
\begin{equation}\label{asymfree_def}
    \avg{p_1(x_{N,i(1)})p_2(x_{N,i(2)})\cdots p_m(x_{N,i(m)})} \rightarrow 0~,
\end{equation}
as $N \rightarrow \infty$ for alternating combinations of the variables (as in the definition of freeness \eqref{free_def}). Since in this paper we shall mostly deal with Hamiltonians of quantum systems with a fixed dimension $N$, we will be interested in the asymptotic freeness of some suitable quantum operators in the limit of large $N$. In the following, we shall often omit the term `asymptotic' and just refer to the variables as free when the limit $N \rightarrow \infty$ is considered. 

\textbf{Example 1.}
One of the canonical examples of asymptotically free variables is large deterministic matrices whose eigenvectors are related by a random rotation. Concretely, consider two sequences of deterministic $N \times N$ matrices, $\BA_N$ and $\BB_N$ having converging DOS in the limit $N \rightarrow \infty$, and the map\footnote{Unless specified otherwise, this is the map we shall use in the rest of the paper. We shall also often denote the map $\varphi(.)$ with angular brackets $\avg{.}$ for convenience, even in cases where no ensemble average needs to be performed. This is sometimes referred in the literature as the \textit{normalised expected trace} (NET), and we shall often use this nomenclature in the following. }  
\begin{equation} \label{trace_map}
    \varphi(\mathcal{O}) = \lim_{N \rightarrow \infty} \varphi_{N}(\mathcal{O})\,,~~~\text{where}~~\varphi_{N}(\mathcal{O}) = \frac{1}{N}\mathbb{E}\left( \text{Tr} ~\mathcal{O}\right)~.
\end{equation}
Then the matrices, $\BA_N$ and $U_N \BB_N U_N^\dagger$, in the limit $N \rightarrow \infty$, are free from each other with respect to the above map, where $U$ is a Haar unitary random matrix \cite{mingo2017free, speicher2019lecture}. 

\textbf{Example 2.} Another example of the asymptotically free variables comes from the following theorem. 
Consider $m$ independent $N\times N$ Wigner matrices\footnote{In mathematics literature, the \textit{Wigner matrices} ($W$) are usually defined as the self-adjoint matrices having independent centred random variables entries on the upper triangle. The diagonal entries are $N$ independent centred real random variables ($W_{ii}=x_d$), while non-diagonal elements are drawn from independent centred real or complex random variables $W_{ij}=x_{od}, i<j$. Furthermore, all the moments of the $x_d$ and $x_{od}$ are assumed to be finite, i.e., there exists a $c_p>0, \forall p \in \mathbb{N}$ such that $\mee |x_d|^p+\mee |x_{od}|^p \leq c_p$. } $W_1, W_2, \cdots W_m$ having entries drawn from distributions with bounded moments of all orders, and construct the rescaled matrices $X_i=W_i/\sqrt{N}$, $i=1,2, \cdots m$. Then the matrices $X_j$ are asymptotically free with respect to the canonical map  \eqref{trace_map}. 

For details of the proofs of the above statements, and further examples, we refer to \cite{voiculescu1992free, nica2006lectures, mingo2017free, speicher2019lecture}. 

\subsection{Cauchy and $R$-transforms}
Now we briefly review the definitions and important properties of some of the key quantities that will be used in the next sections 
to obtain the DOS of quantum Hamiltonians.
We start by considering the \textit{Cauchy transform} of the DOS ($\rho_{\mathcal{O}} (\lambda)$) of an operator $\mcO$ (which is the trace of the resolvent of the operator),\footnote{For a given Hermitian matrix $\BA \in \mrr^{N \times N}$ with eigenvalues $a_i$, the Cauchy transform of its empirical eigenvalue distribution is given by
\begin{equation}
    G^N_\BA(z) = \frac{1}{N} \text{Tr} \Bitd{ \Big(z \BI -\BA\Big)^{-1}} = \frac{1}{N} \sum_{i=1}^N \frac{1}{z-a_i}~.
\end{equation} 
Consider a probability measure, say, $\rho$, which has a compact support on the real line.  Then  there exists a $*$-probability space (given by the algebra $\CA$, and the linear map $\varphi$) such that there is a self-adjoint random variable $x \in \CA$, with $\rho_x=\rho$.}
\begin{equation}\label{Cauchy_O}
	G_{\mathcal{O}}(z) = \int \frac{\rho_{\mathcal{O}}(\lambda)}{z-\lambda} ~ \text{d}\lambda~.
\end{equation}
The integral above is over the support of the distribution $\rho_{\mathcal{O}} (\lambda)$ on the real line, and $G_{\mathcal{O}}(z)$ is analytic in the upper half of the complex plane, i.e., for all $z \in \mcc^+:=\{z \in \mcc|\text{Im}(z)>0\}$. When the Cauchy transform has more than one branch in the complex plane, we choose the one that has $\text{Im}(G_{\mathcal{O}}(z))<0$ whenever $\text{Im}(z)>0$. Also, we consider the solution which vanishes at infinity, i.e., $G_{\mathcal{O}}(z) \rightarrow 0$ as $|z| \rightarrow \infty$.

In mathematics literature, it is also common to define the negative of the Cauchy transform, known as the \textit{Stieltjes transform}. One of the most widely used approaches for numerically estimating the Cauchy transform is to employ the Lanczos algorithm, where one finds a continued fraction approximation of the resolvent operator. To overcome the slow convergence and numerical instability of the Lanczos algorithm-based method, in a recent work \cite{ameli2025spectral}, two other alternative methods, based on the Jacobi and Chebyshev polynomials, were presented for numerically approximating the Stieltjes transform of a Hermitian operator. 

Note that, for a self-adjoint operator $\mcO$ of rank $N$, the Cauchy transform can be thought of as its moment generating function, i.e., in terms of the moments of the operator, $\varphi(\mcO^k)$, it can be expanded as
\begin{equation}
    G_\mathcal{O}(z)=\sum_{k=0}^{\infty} \frac{\varphi(\mathcal{O}^k)}{z^{k+1}}~,~\text{for all}~z \in \mcc^+~,~~ \text{with}~~\varphi(\mathcal{O}^k)= \frac{1}{N} \mathbb{E}(\text{Tr}(\mcO^k))= \int \rho_\mathcal{O} (\lambda) \lambda^k \ted \lambda~.
\end{equation}
In the final expression on the right-hand side above, we have written the moments (for every $k \in \mnn$) in terms of the DOS of the operator, to which the eigenvalue density of the operator $\mcO$ converges in the $N \rightarrow \infty$ limit. We shall assume that for the Hamiltonians we consider in the following, the DOS has a compact support on the real line, so that it is completely determined by the moments of the Hamiltonian $\langle H^n \rangle$, which in turn, can be obtained from $G_{H}(z)$ in the neighbourhood of $|z| \rightarrow \infty$. 

Next, we define the \textit{$R$-transform} associated with the DOS of the operator $\mathcal{O}$, constructed from the functional inverse of $G_{\mathcal{O}}(z) $. This is given by the following relation \cite{voiculescu1986addition, speicher2019lecture, mingo2017free}
\begin{equation}\label{B_and_R1}
	G_{\mathcal{O}}^{-1}(z) \equiv \mathcal{B}_{\mathcal{O}}(z) = \frac{1}{z} + R_{\mathcal{O}}(z)\,,
\end{equation}
where the meromorphic function $\mathcal{B}_{\mathcal{O}}(z)$ is the functional inverse of the Cauchy transform and has a single pole at $z=0$, so that $R_{\mathcal{O}}(z)$ is analytic in a neighbourhood of $z=0$. 
The $R$-transform obtained from the DOS  of an operator is the generator of its \textit{free cumulants} ($\kappa_j(\mcO)$) \cite{nica2006lectures, mingo2017free}, i.e., we have the expansion,
\begin{align}\label{free_cumuluants}
    R_\mcO(z) = \sum_{j=1}^{\infty} \kappa_j(\mcO) z^{j-1}\,.
\end{align}
Since we have assumed that the DOS is compactly supported on the real line, the free cumulants are bounded, and the series expansion for the $R$-transform converges for sufficiently small values of $|z|$.\footnote{For more detail on the analytical properties of the $R$-transform for a generic distribution, we refer to standard references on the subject, e.g.,  \cite{nica2006lectures,speicher2019lecture}.}

To understand the significance of the free cumulants and $R_\mcO(z)$, recall that, in classical probability theory, the classical cumulants linearise the convolution of two probability distributions, and the logarithm of the Fourier transform of the distribution function is the generating function of these classical cumulants. In analogy with these facts, in free probability theory, the free cumulants linearise the process of additive convolution of two free operators, i.e., if $\BA$ and $\BB$ are free, then $\kappa_n(\BC)=\kappa_n(\BA)+\kappa_n(\BB)$, furthermore, the $R$-transform, which generates these cumulants, can be thought of as the analogue of the log-Fourier transform \cite{nica2006lectures}.

To see quantum mechanical implications of the preceding definitions, consider taking the operator $\mathcal{O}$ to be the Hamiltonian ($H$) of a quantum mechanical system drawn from some random matrix ensemble, then we can relate the Cauchy transform of $H$ to the time evolution operator generated by the Hamiltonian. To see this, start from the trace of the ensemble-averaged time evolution operator (i.e., the NET of the usual time evolution operator),
\begin{equation}
    \mathcal{U}_H(t) = \frac{1}{N}\mathbb{E}\big(\text{Tr} ~e^{-i H t}\big) = \avg{e^{-i H t}}\,.
\end{equation}
It can be straightforwardly checked that this is related to the DOS of the Hamiltonian by a Fourier transform 
\begin{equation}\label{dos_FT_U}
    \rho_{H}(\lambda)= \frac{1}{2 \pi}\int_{-\infty}^{\infty} \text{d}t~e^{i t \lambda}~\mathcal{U}_H(t)\,.
\end{equation}
Now using this relation and the definition of the Cauchy transform \eqref{Cauchy_O}, we see that $G_H(z)$ is related to the averaged time evolution operator as\footnote{Here we have written this expression for $\Im z<0$ to ensure convergence. One can, of course, write an analogous, properly convergent, expression for $\Im z>0$.} (see e.g., \cite{BREZIN1996697})
\begin{equation}\label{G_evolution}
    G_{H}(z) = i  \int_{0}^{\infty} ~ \text{d}t ~ \mathcal{U}_H(-t) ~e^{-itz}\,~,~\Im z<0~.
\end{equation}
Although we have written these relations for the Hamiltonian here, similar relations are valid for any self-adjoint operator.

Finally, we note that, substituting $z \rightarrow \CB_\CO(z)$ in eq.  \eqref{G_evolution} written for an operator $\CO$, we have the following relation involving the evolution operator and the $R$-transform of its eigenvalue distribution
\begin{equation}\label{B_evolution}
    z = i  \int_{0}^{\infty} ~ \text{d}t ~ \mathcal{U}_\CO(-t) ~e^{-it\CB_\CO(z)} = i  \int_{0}^{\infty} ~ \text{d}t ~ \mathcal{U}_\CO(-t) ~\exp\Big[{-it\Big(R_{\CO}(z)+\frac{1}{z}\Big)}\Big]\,,~~\Im z<0~.
\end{equation}

\subsection{Free additive convolution}\label{additive_conv}
Consider two operators $\BA$ and $\BB$ having compact DOS which are asymptotically free, then one can show that the DOS of the sum $\BC=\BA +\BB$ depends only on the DOS of the individual operators and $\rho_\BC$ is a probability measure having compact support over the real line. The procedure of obtaining the DOS of $\BC$, which is the analogue of the classical convolution in the classical probability theory of commutative random variables, is known as the \textit{free additive convolution},  and usually denoted as $\rho_\BA \boxplus \rho_\BB = \rho_\BC$. Since in this manuscript we mainly consider the DOS of sums of two operators, not multiplication or other more complex operations, we will often refer to the free additive convolution as just free convolution for brevity. For a free probabilistic description of the eigenvalue distribution of the product of two non-commutative free random variables, see \cite{voiculescu1987multiplication,nica2006lectures, Potters}. The particular analytical tool which is used for this purpose is known as the $S$-transform, and can be obtained from the knowledge of the $R$-transform. For the use of free probability to find the spectral distributions of more complicated combinations of non-commutative random variables,  apart from sums and products of them, see e.g., \cite{nica1998commutators}.

The method for performing the free additive convolution uses the property of the $R$-transforms that for free variables it is additive, i.e., the $R$-transform of the DOS of the sum $\BC=\BA +\BB$ is the sum of the $R$-transforms of the  DOS of the individual operators $\BA$ and $\BB$, if they are free \cite{voiculescu1992free, speicher2019lecture, nica2006lectures, voiculescu1986addition, speicher1993free }.\footnote{For calculation of the DOS and the eigenvalue correlation for sum of two random matrices or a random and deterministic matrix based on diagrammatic method, see \cite{brezin1, brezin2, zee1996law}. See the last reference for a derivation of the additivity of the $R$-transforms for random matrices using a diagrammatic method. See also \cite{zinnjustin1999adding} for an alternative proof based on the Harish Chandra-Itzykson-Zuber integration formula \cite{Potters}.} From the 
$R$-transform of the operator $\BC$, one can use the procedure of obtaining the formula \eqref{B_and_R1} in a backward manner to compute the Cauchy transform of the DOS $\rho_{\BC}$ of $\BC$, and hence, the DOS itself by employing the \textit{Stieltjes inversion formula}  since $\rho_\BC$ has a compact support on the real line. According to the Stieltjes inversion formula, the continuous part of the DOS ($\rho_\CO(\lambda)$) of an operator $\CO$ is determined by the non-zero imaginary part of the Cauchy transform on the real line,
\begin{equation}\label{dos_from_G}
	\rho_{\CO}(\lambda) = -\frac{1}{\pi}\lim_{\epsilon \rightarrow 0^+} (\text{Im} ~ G_{\CO}(\lambda+ i \epsilon))\,.
\end{equation}
Furthermore, if $G_{\CO}(z)$ has poles on the real axis, the DOS contains Dirac masses with magnitude at each point being given by the residue of $G_{\CO}(z)$ at that pole.
The function $\CB(z)$ being inverse of the Cauchy transformation, can have multiple branches, so that during computation of the $R$-transform one should choose the branch which satisfies the condition $R(0)=0$.

There are some well-known examples in the literature where one can obtain an analytical expression for the DOS of the operator $\BC$ by following the steps of the free additive convolution. These examples include, e.g.,  the cases of the addition of two free spin 1/2 or spin 1 operators \cite{speicher2019lecture, Camargo:2025zxr}, as well as the sum of two operators having constant DOS \cite{Potters}.
However, in general, for given DOS of the operators $\BA$ and $\BB$, analytically performing the free convolution procedure can be quite difficult since it involves inverting a polynomial equation for the Cauchy transform $G_\BC(z)$. A very useful method of numerically obtaining the DOS by performing the free convolution is to use the subordination formulas (see below) for the Cauchy transform for the  DOS of $\BC$ \cite{speicher2019lecture}. In this context, we refer to \cite{olver} for a numerical recipe of performing the free convolution, which seems to work very well when the DOS of the individual operators have connected support. For further examples of numerical free convolution involving spin operators, see  \cite{Camargo:2025zxr, Jahnke:2025exd, Pollock:2025acf} and \cite{capitaine2016spectrum} for the application of the free probability in understanding the spectral properties of deformed random matrices. We also refer to the recent work \cite{cortinovis2025computing}, which proposed an algorithm for computing the free additive convolution of two given probability measures with compact supports (and continuous without Dirac masses) whose free convolution has a square-root behaviour at the edges. 

\paragraph{Subordination formulas.}
Since $\BA$ and $\BB$ are free, their $ R$-transforms are additive, i.e., $R_{\BC}=R_{\BA}+R_{\BB}$ \cite{mingo2017free, voiculescu1992free, nica2006lectures, Potters}, hence one can write, 
\begin{equation}
	\mathcal{B}_{\BC}(z) = R_{\BC}(z)+\frac{1}{z}\,~~ \Rightarrow ~~ \mathcal{B}_{\BA}(z) = \mathcal{B}_{\BC}(z)- R_{\BB}(z)\,.
\end{equation}
Substituting $z \rightarrow G_{\BC}(z)$ in the argument of the expression above, and using the fact that $G_\BC$ and $\mathcal{B}_\BC$ are functional inverses to each other, we have 
\begin{equation}
	\mathcal{B}_\BA\big(G_{\BC}(z)\big) = z- R_{\BB} \big(G_{\BC}(z)\big)\,.
\end{equation}
Now applying $G_{\BA}$ on both sides of this expression, we get  the following self-consistency equation 
for the Cauchy transform of the $G_{\BC}(z)$,
\begin{equation}\label{subor_A}
	G_{\BC}(z) = G_{\BA}\big(z-R_{\BB}\big(G_{\BC}(z)\big) \big)\,.
\end{equation}
Similarly, we can derive the analogous relation,
\begin{equation}\label{subor_B}
	G_{\BC}(z) = G_{\BB}\big(z-R_{\BA}\big(G_{\BC}(z)\big) \big)\,.
\end{equation}
Knowing the expressions for the Cauchy transforms $G_{\BA}(z)$ and $G_{\BB}(z)$ from the given DOS of these operators, we can, in principle, use one of the last two formulas to solve for $G_{\BC}(z)$ to subsequently get the DOS of the operator $\BC$ using the inversion formula. 
However, for given DOS of the two operators $\BA$ and $\BB$, in most cases of physical interest, the expressions for the corresponding Cauchy transforms and the $R$-transforms are quite complicated, and later is difficult to obtain analytically due to the functional inverses of $G(z)$ involved in these formulas. Hence, except for some simple situations, the subordination formula can not be solved exactly for $G_{\BC}(z)$. One way we can make some progress is by transforming to the Fourier representation of the DOS and subsequently using a series expansion. 
In Section \ref{dos_perturb} we shall follow this approach to develop a perturbation series for $G_\BC(z)$ from these subordination formulas to obtain an approximate DOS of $\BC$. 

Before moving on, we note one further important point regarding the subordination formulas above.  Let us define the \textit{subordination functions }
\begin{equation}
    \omega_{\BA}(z):=z-R_{\BB}\big(G_{\BC}(z)\big)~,~~\text{and}~~\omega_{\BB}(z):=z-R_{\BA}\big(G_{\BC}(z)\big)~,
\end{equation}
so that the relations in \eqref{subor_A} and \eqref{subor_B} reduce to $G_{\BC}(z)=G_{\BA}(\omega_{\BA}(z))=G_{\BB}(\omega_{\BB}(z))$. 
Furthermore, defining the functions 
\begin{equation}
    \CH_\BA(z):= \frac{1}{G_{\BA}(z)}-z ~,~~\text{and}~~\CH_\BB(z):= \frac{1}{G_{\BB}(z)}-z~,
\end{equation}
one can derive a fixed point equation for the subordination functions (see e.g., \cite{speicher2019lecture, mingo2017free} for the proof) 
\begin{equation}\label{subor_omega}
    \omega_{\BA}(z)=z+\CH_{\BB} \Big[z+\CH_\BA(\omega_{\BA}(z))\Big]~
\end{equation}
which only involves the (variants of) Cauchy transforms, not the $R$-transforms of the DOS. An important advantage of this reformulation of the relations \eqref{subor_A} and \eqref{subor_B} is that, since the subordination functions can be extended to the entire upper-half complex plane ($\mcc^+$),  the relation  \eqref{subor_omega} is analytically well-behaved compared to the $R$-transform relation, as $R(z)$ can not be extended analytically to all of $\mcc^-$. Furthermore, for the given DOS of the operators $\BA$ and $\BB$, this equation can be solved numerically by starting, for a fixed value of $z \in \mcc^+$, with an initial guess for $\omega_{\BA}(z)$, and iteratively updating this ansatz until it converges.  See \cite{Camargo:2025zxr, Pollock:2025acf} for recent works in the physics literature where this method has been used to compute DOS, including the cases of addition of spin operators. 

\section{Density of states of quantum systems from free additive convolution}\label{sec_dos_quantum}

As we have discussed in the Introduction, our goal is to provide an overview of the computation of the DOS of realistic quantum systems using the tools of free probability.  
In this section, we review several examples of situations where the DOS is calculated by writing the Hamiltonian as $H= \BA + \BB$, and assuming that the component operators $\BA$ and $\BB$ are free with respect to each other. Given the operators $\BA$ and $\BB$ for the Hamiltonian of a realistic quantum system or a random matrix Hamiltonian, as an effective strategy of checking whether these operators are asymptotically free or not, one can first compute the mixed cumulants $\kappa_n(\BA, \BB)$ between $\BA$ and $\BB$ and see if these vanish as the dimension of the matrices $N$ is increased. If these two operators are asymptotically free, one would expect that $\kappa_n(\BA, \BB) \rightarrow 0, \forall n$ as $N \rightarrow \infty$.\footnote{Of course, the exact way these mixed free cumulants vanish depends on the nature of the system under consideration, and there remains much to be explored on the scaling of the free cumulants in realistic quantum systems.}

In this section, we also discuss the matching of the free probability approximation to the DOS with the numerical result obtained from exact diagonalisation.  
In the next section, we shall develop an analytical approximation method to approach this problem. 

\subsection{Density of states of a class of local random matrix Hamiltonians}\label{ssec_LRMT}

As a first example, let us consider a class of local random matrix Hamiltonians, which were recently studied in \cite{Pollock:2025acf} to investigate the transport of energy in a quantum system with spatial locality. As we shall see, for the case of few-body Hamiltonians, one can map the problem of computing DOS to the problem of computing the single-particle DOS in a nearest-neighbour hopping Hamiltonian. In some specific few-body interacting cases, as well, the DOS can be computed exactly.
Alternatively, one can employ free probability directly to compute the DOS, using the freeness between the components—a fact that can be proven analytically, e.g., for the case of a three-site nearest-neighbour chain under either open or periodic boundary conditions. 

The basic building block of this Hamiltonian is a nearest-neighbour interaction term coupling sites levelled by $j$ and $j+1$, 
which can be written as $h_j=U_j\Lambda U_j^{\dagger} \otimes I_{\bar{j}}$. Here the matrices $\Lambda$ and $U_j$ are of dimension $q^2 \times q^2$, while $ I_{\bar{j}}$ is an identity matrix 
which acts on all sites apart from $j$th and $(j+1)$th. The matrices $U_j$ are drawn from independent unitary Haar random distributions, while the matrix $\Lambda$ is chosen in such a way that $\Lambda^2=1$ and $\langle \Lambda \rangle=0$.  Note that these conditions imply $\avg{h_i}=0$ and $h_i^2=I$. 

The Hamiltonian considered in \cite{Pollock:2025acf} has the form $H=\sum_{j} h_j$, where the sum is over $\eta$ Haar-Ising bonds, i.e., $\eta-1$, $\eta$ being the number of Haar-Ising terms. For example, in the simplest non-trivial case with open boundary conditions, the system consists of three sites and two Haar-Ising bonds. The  Hamiltonian in that case can be written as $H=h_0+h_1$, so that $I$ here is a $q \times q$ identity matrix, and each of the $h_i$s is therefore a $q^3 \times q^3$ matrix.  We first consider this three-site system to illustrate the computation of the DOS. 

Before proceeding further, we first state an important result established in \cite{Pollock:2025acf} using the techniques of Weingarten calculus \cite{collins2022weingarten}.  The result asserts that in the large $q$ limit $q \rightarrow \infty$ the alternating moments of the Haar-Ising terms of the form $\avg{(h_0h_1)^m}$ vanish for all $m$, i.e., 
\begin{equation}\label{mixed_h01}
    \avg{(h_0h_1)^m}=0~~~\text{with}~~~q\rightarrow\infty~.
\end{equation}
We refer to \cite{Pollock:2025acf} for the derivation of this result (see also \cite{Fava:2023pac, Collins:2022klx, charlesworth2021matrix}). A straightforward consequence of this result (along with the fact that $h_i^n-\avg{h_i^n}I=0$ for $n$ even, and equal to $h_i$ for $n$ odd) is that Haar-Ising interacting terms are asymptotically free in the large $q$ limit (see the definition \eqref{free_def}). 

First, let us compute the DOS by directly computing the moments $\avg{H^n}$. It is easy to see that for $H=h_0+h_1$, the $n$th moment is a sum of all possible strings of the terms $h_0$, and $h_1$, i.e., $\avg{H^n}=\sum_{b} \avg{h_{b_1}h_{b_2}h_{b_3}\cdots h_{b_n}}$. Now, if the binary string $b$ is such that it can be fully reduced using the relation $h_i^2=1$, then that term would be unity. On the other hand, if it is not possible to fully reduce the string, then eq. \eqref{mixed_h01} implies that it would be zero in the limit $q \rightarrow \infty$. Hence, to compute the moments, we need to count the number of fully reducible strings, which in turn can be done by mapping the problem to the problem of counting the trajectories of a particle hopping in an infinite one-dimensional chain, where the fully reducible string would correspond to the case when the particle hops back to its initial position. 

This can be done conveniently by introducing the following nearest neighbour hopping Hamiltonian 
\begin{equation}\label{H_hopping}
    \Delta = \sum_{x \in \mathbb{Z}} \ket{x-1}\bra{x} + \ket{x+1}\bra{x}~,
\end{equation}
and noting that the moments of the local Haar-Ising Hamiltonian can be written as $\avg{H^n}=\avg{0|\Delta^n|0}$ since this is the amplitude of the particle returning to the initial position, and hence the number of completely reducible strings. Note that in the notation used in \eqref{H_hopping}, $\ket{x}$ denotes a single particle sitting at a site of position $x$, hence $\ket{0}$ denotes a particle sitting at the origin. Also note that, when written in this basis, the Hamiltonian $\Delta$ has a tridiagonal form with constant elements (This type of matrix will be very useful and appear frequently in the later part of this section, as well as in the following sections).

Using the Hamiltonian \eqref{H_hopping} it can seen that for odd values of $n$, $\avg{H^n}$ vanishes, while for even values of $n=2k$
the moments are given by the binomial coefficients $\avg{H^{2k}}=(2k)!/(k!)^2$. As it is well-known, these are the moments of the \textit{arcsine distribution}\footnote{The arcsine distribution is a special case of the \textit{Kesten–McKay distribution} \cite{kesten1959symmetric, mckay1981expected} written as 
\begin{equation}\label{KM_dist}
    \rho^{KM}(\lambda)=\frac{\eta \sqrt{4(\eta-1)-\lambda^2}}{2 \pi ~(\eta^2-\lambda^2)}~,~~~\eta \geq 2~,
\end{equation}
which has a support $|\lambda| \leq 2 \sqrt{\eta-1}$ in the real line. The $\eta$-fold free additive convolution of the Bernoulli distribution in \eqref{Bernoulli_dist} (with $\eta \in \mrr$, and $\eta \geq 2$) gives the Kesten-McKay distribution \cite{nica2006lectures, speicher2019lecture}.} 
\begin{equation}\label{arcsine}
    \rho_{\text{as}}(\lambda)=\frac{1}{\pi \sqrt{4-\lambda^2}}~,~~\text{for}~~|\lambda|<2
\end{equation}
which is therefore the DOS of the local random matrix Hamiltonian under consideration with two Haar-Ising terms in the limit $q \rightarrow \infty$. 

This conclusion can also be reached using the free probability theory directly. To this end, we just note that for the Hamiltonian $H=\sum_{j=1,2} h_j$ with $h_1$ and $h_2$ being asymptotically free from each other in the large $q$ limit, the DOS of $H$ is just the free additive convolution of the DOS of $h_1$ and $h_2$. From our construction, both $h_i$ have the Bernoulli distribution as the DOS 
\begin{equation}\label{Bernoulli_dist}
    \rho_{h_i}(\lambda)=\frac{1}{2}\Big(\delta(\lambda+1)+\delta(\lambda-1)\Big)~,
\end{equation} 
and hence, as is well known in the free probability literature \cite{nica2006lectures, speicher2019lecture}, their free convolution gives the arcsine distribution \eqref{arcsine} as the DOS of $H$. 

We now proceed to discuss the case with more than two non-commutating Haar-Ising terms, where the Hamiltonian reads, $H=\sum_{j=0}^{\eta-1} h_j$, with all the terms of the Hamiltonian being mutually non-commutating, i.e., $[h_i,h_j] \neq 0, \forall i,j, i \neq j$, and the exact structure of the chain, i.e., on which site the local terms act, is to be specified (see Fig. 4 of \cite{Pollock:2025acf} for different arrangements). It was pointed out in\cite{Pollock:2025acf} for Hamiltonians containing more than two non-commuting Haar-Ising term, the mixed moments (i.e., the irreducible words) of $h_i$s also vanish in the large $q$ if the non-commutating terms are arranged in such a way that all the bonds overlap at least in a single site (see also \cite{Collins:2022klx, charlesworth2021matrix}).\footnote{For an arrangement where three Haar-Ising interaction terms are arranged in a three-site periodic chain, it was observed in \cite{Pollock:2025acf} that the irreducible words of $h_i$ also vanish in this case in the $q \rightarrow \infty$ limit (recall that the $h_i$ terms are pairwise free in the large $q$ limit).  We thank Klée Pollock and Jonathon Riddell for useful correspondence on this point.}
Therefore, analogous to the case of two interacting terms discussed above, in these cases as well, the moments of the Hamiltonian can be realised as a single-particle hopping problem on a Bethe lattice \cite{ostilli2012cayley} with coordination number $\eta$. Ref \cite{Pollock:2025acf}  used this mapping to compute the DOS of a Hamiltonian where three $h_i$ bonds are arranged on three sites in the presence of periodic boundary conditions, and it was shown that it provides a good approximation to the exact numerical DOS for this problem.  

Let us now discuss the use of free probabilistic techniques to compute the DOS of the local random matrix Hamiltonian with more than two Haar-Ising interacting terms. The discussion in the previous paragraph indicates that, in this case, as well, the non-commutating Haar-Ising terms are asymptotically free in the limit $q \rightarrow \infty$. 
The philosophy is as explained at the beginning of this section: we write the total Hamiltonian as the sum of two operators $\BA$ and $\BB$ with known DOS, and compute the DOS of the Hamiltonian using the free additive convolution by assuming that these two operators are mutually free. 
However, when there is more than one non-commutating term in the Hamiltonian, there are several choices for the combination one can use to perform the free convolution. E.g., when $H=\sum_{i=1}^{4} h_i$, one can take (I) $\BA=h_1+h_3$ and $\BB=h_2+h_4$, or (II) $\BA=h_1+h_2$ and $\BB=h_3+h_4$. Notice the important difference between the two cases for the local Haar-Ising interacting Hamiltonian we are considering: for the first case, the components of both the operators are mutually commutating, hence a classical convolution has to be used to compute the DOS of $\BA$ and $\BB$ in this case, which can be subsequently free convolved to obtain the DOS of $H$.   Whereas, in the second case (II), since $h_1$ and $h_2$ are non-commutating, one has to use free convolution to obtain DOS of $\BA$ (also for $\BB$), which then needs to be further free convolved to obtain the DOS of the Hamiltonian. It was verified in \cite{Pollock:2025acf} that for the case four Haar-Ising terms with periodic boundary conditions, the DOS obtained in case I matches with the numerical result, whereas for case II it does not. 

Before proceeding, we would like to note the following points. 
1. The precise reason why Case I produces the correct DOS, while Case II does not, is not always clear (see the discussion below on the isotropic entanglement in this context), and furthermore, the splitting $H=H_{\text{even}}+H_{\text{odd}}$ and subsequent use of free convolution does not always produce the correct DOS,  as was pointed out in \cite{Pollock:2025acf} by considering a chain with six interacting Haar-Ising terms.\footnote{The chain with four Harr-Ising terms with periodic boundary conditions mentioned above is a special case in this regard, where there exists a decomposition $H=\BA+\BB$ with $\BA$ being asymptotically free from $\BB$. This is because the operator $h_1$  is asymptotically free from both $h_2$ and $h_4$ (and commutes with $h_3$). In general, for longer chains, this is not true, and one needs to employ the so-called $\epsilon$-free probability to understand these cases \cite{Pollock:2025uem}. We thank Klée Pollock for explaining this point to us.} 2. For chains with a large number of interacting terms, the DOS converges to a Gaussian distribution \cite{Pollock:2025acf, Pollock:2025uem}, for both small $q$ and the limit of large $q$, as long as there are many commutating terms. Note that the case of chains with small values of $q$ is opposite to the free probability limit $q\rightarrow \infty$, which corresponds to chains having a large number of degrees of freedom at each site.

\paragraph{Isotropic entanglement.}
We note that this idea of separating the even and odd terms of a Hamiltonian with only nearest neighbour interaction and use free probability to compute the DOS of $H$ by first using classical convolution to obtain the DOS of sum of commutating even or odd terms separately is analogous to a proposal of obtaining the DOS of quantum spin systems with generic local interactions developed in \cite{movassagh2010isotropic, PhysRevLett.107.097205} by matching the fourth moment of the DOS, which was referred to as the \textit{isotropic entanglement}. Here, we briefly review this idea in connection with the discussion we have so far in this section, while in Section \ref{sec_sum_AB} we provide an overview of the fourth moment matching method in general terms.  

Consider a one-dimensional chain with
$\eta$ generically interacting $q$-dimensional qudits, whose  Hamiltonian is written as 
\begin{equation}\label{spin_hamiltonian}
    H = \sum_{j=1}^{\eta-1} I_{q^{j-1}} \otimes h_{j,\cdots, j+\xi-1} \otimes I_{q^{\eta-j-(\xi-1)}}~,
\end{equation}
where, similar to the case of the local random matrix model considered at the beginning of this section, $h_{j,\cdots, j+\xi-1}$ are $q^\xi \times q^\xi$ random matrices, with $\xi=2$ corresponding to the case nearest neighbour interaction, which we also consider below. One can obtain the eigenvalue spectrum of any commuting subset of the total Hamiltonian (e.g., sum of the terms with $j$ even or sum with $j$ odd in \eqref{spin_hamiltonian}) by diagonalising, and to obtain the DOS of $H=\BA+\BB$, where $\BA=\sum_{j=1,3,5, \cdots} I \otimes h_{j,j+1}\otimes I$ and $\BB=\sum_{j=2,4,6, \cdots} I \otimes h_{j,j+1}\otimes I$, one needs to calculate a parameter $p$ which provides an interpolation between the fourth moment of the classical convolution and free convolution \cite{movassagh2010isotropic}.

\paragraph{For generic operators.} The arguments presented in this section so far can be utilised to see that for a Hamiltonian of the form $H=\BA+\BB$
where both the operators $\BA$ and $\BB$ are traceless, and both square to unity (i.e., $\BA^2=\BB^2=1$), and are free from each other, the DOS of the Hamiltonian is given by the arcsine distribution. To see this, we consider the moments of the Hamiltonian 
$\avg{H^n}=\avg{(\BA+\BB)^n}$. Using the fact that $\BA^2=\BB^2=1$ and the operators are free and traceless, it can be seen that the moments are the sum of fully reducible strings, which, as in the case of the Haar-Ising model, gives the binomial coefficients $(2k)!/(k!)^2$ as moments for $n=2k$, while all the odd moments vanish. E.g., the fourth-order moment can be written explicitly as 
\begin{equation}
    \avg{H^4} = 6+ 2 ~\avg{\BA\BB\BA\BB}~.
\end{equation}
The second term in the RHS of the above expression vanishes since the component operators are free and traceless, and the moment reduces to the required binomial coefficient. 

Examples of such cases include spin 1/2 $X_j$ and $Y_j$ operators constructed from the tensor product of Pauli matrices, which are usually constructed to act, say, on the $j$th site of a chain of length $L$.  Of course, these operators being deterministic are not, in general, free from each other. In fact, two such operators acting on different sites of the chain with open boundary conditions commute with each other. However, one can produce asymptotically free operators from them by performing a suitable unitary transform of one of these operators. Examples of such unitary transforms include the Haar random unitary operator \cite{nica2006lectures, speicher2019lecture, Potters} and also, as was recently shown, the time evolution operator generated by chaotic Hamiltonians \cite{Camargo:2025zxr, Chen:2024free}.

\subsection{Density of states of disordered quantum systems.}\label{sec_dos_anderson}

\textbf{Anderson Hamiltonian (with on-site disorder).}
The next example we consider is the Anderson Hamiltonian, which describes the hopping of an electron in a $d$-dimensional lattice in the presence of disorder  \cite{Andersonlocal}. 
Let us consider the case of a one-dimensional chain having nearest-neighbour interaction, where the Hamiltonian takes the following form
\begin{equation}\label{H_anderson}
    H = \begin{pmatrix}
h_1 & J   &       &        &        \\
J   & h_2 & J     &        &        \\
    & J   & \ddots & \ddots &        \\
    &     & \ddots & h_{N-1} & J     \\
    &     &        & J      & h_N
\end{pmatrix}~,~~~\text{i.e.}~~H_{ij}=h_i \delta_{ij}+ J_{ij}~.
\end{equation}
Here, the diagonal elements $h_i$ are i.i.d. random variables drawn from a given probability distribution function $p_h(x)$, and first we consider the situation with $J_{ij}=J M_{ij}$, with $J$ being a constant and $M$ being essentially the adjacency matrix of a one-dimensional chain. The free probability theory was used in \cite{Chenprl} (see also \cite{neu1995rigorous} and section \ref{Anderson_free} below) to explain the behaviour of the DOS of this Hamiltonian for different choices of the distribution $p_h(x)$ and the constant $J$. 

To apply the free probability theory to compute the DOS of a Hamiltonian of the form \eqref{H_anderson}, the first point we notice is that for arbitrary values of the constant $J$ and the probability distribution $p_h(x)$, there is no unique choice for component operators $\BA$ and $\BB$. To this end, one takes some `natural' choices, and sees which one gives a  better approximation to the DOS compared with the numerically obtained ones.\footnote{In cases where there is a natural small parameter, $\alpha$, with respect to which one can set up a perturbation series, in some sense, there is a `natural' choice for these operators. See the method described in the next section (section \ref{dos_perturb}) and the discussion therein. If one is given the value of the constant  $J$, and the choice of the distribution $p_h(x)$, then, as we shall see, one can make a suitable choice of these operators.}

To this end, two different choices were considered in \cite{Chenprl}: I. In the first choice, the operator $\BA$ was taken to be a diagonal matrix whose elements are i.i.d. random variables drawn from $p_h(x)$, so that the matrix $\BB$ is a tridiagonal matrix with zero diagonal elements and all the off-diagonal elements being the constant $J$ i.e., it is the adjacency matrix of a one-dimensional chain multiplied by the constant $J$. II. The second choice is where both the operators $\BA$ and $\BB$ are constructed from the $2 \times 2$ matrices of the form \cite{Chenprl}
\begin{equation}\label{H_2}
    H =\left(
		\begin{array}{ccc}
			h_1 & J\\
			J  & 0
		\end{array}
		\right)~. 
\end{equation}
For the scheme I, it was shown in \cite{Chenprl} by choosing the distribution $p_h(x)$ to be either Gaussian or Wigner semicircle, the DOS predicted by the free probability theory matches very well with the numerical results for all range of the `magnitude' of the noise\footnote{The magnitude of the noise is quantified by the ratio of (denoted by $\sigma$) the variance of the distribution of the matrix $\BA$ in scheme I  with the constant $J$. }. Whereas, for the choice of the operators in scheme II, the quality of matching between the free convolution prediction and the numerical result depends on the magnitude of the noise: if the distribution $p_h(x)$ is a Gaussian with mean $0$ and variance $\sigma^2$, the fitting gets worse as the ratio $\sigma/J$ is increased. 

The difference in predicting the exact distribution using free probability theory for these two schemes can be explained by considering the $n$-th order mixed moments of the component operators $\BA$ and $\BB$ which is of the form $\langle{(\BA \BB)^{n}}\rangle$  and comparing these with those predicted by the free probability theory \cite{nica2006lectures, mingo2017free}. If for the scheme I the match is, say, up to order $n_1$, whereas for the other scheme II it is up to $n_2$, with $n_1>n_2$, then the DOS in scheme I should be closer to the exact DOS than that of scheme II. In \cite{Chenprl}, this was shown exactly to be the case for the schemes described above.\footnote{By taking $\BA$ to be a diagonal matrix with i.i.d. Gaussian variables as elements, and $\BB$ a tridiagonal matrix with elements $\BB_{ij}=J\big(\delta_{i,j+1}+\delta_{i,j-1}\big)$, it can be easily checked by explicit calculation that the first non-zero moment is $\avg{(\BA \BB)^{4}} = 2J^4 \sigma^4$, hence the finite dimensional matrices $\BA$ and $\BB$ here are partially free, up to seventh order in the moments.} Nevertheless, we note that, even though the moment matching gives an explanation of why scheme I works better than scheme II, it does not rule out the possibility of finding a pair of matrices, which, when taken as the operators $\BA$ and $\BB$, approximate the exact DOS of the one-dimensional Anderson model better than the previous two cases.  

The case of the Anderson model in one dimension with nearest-neighbour interaction, which we have discussed so far, was extended in \cite{Welborn} to Hamiltonians with constant long-range interactions and/or defined on two and three-dimensional lattices. Again, it was found that, for these systems with only on-site disorder, the exact DOS obtained numerically can be approximated quite well by using the free probability theory.

\textbf{With off-diagonal disorder.} In addition to the on-site disorder, if one considers the presence of off-diagonal disorder as well, then the DOS obtained from free probability theory was shown to be not a very good approximation to the exact DOS. E.g., the simplest such model with random nearest-neighbour interaction 
\begin{equation}
    H_{ij}=h_i \delta_{ij}+ J_{i}\big(\delta_{i,j+1}+\delta_{i,j-1}\big)~,
\end{equation}
was considered in \cite{Welborn} (where $h_i$ are now random variables), and it was shown that the mixed moments $\langle{(\BA \BB)^{n}}\rangle$ in this case matches with the free probability prediction only up to sixth order, rather eighth order as in the case of constant interaction - thereby justifying the relatively poor matching of the free probability prediction and the exact DOS in the case of disordered interaction.  

Before moving on, we note that, in these examples where the operators $\BA$ or $\BB$ are not random matrices, a priori, there is no justification for assuming that these operators are free.\footnote{There are known examples where two finite-dimensional deterministic matrices are partially free from each other and become asymptotically free in the limit when their dimension goes to infinity (see, e.g., \cite{chen2012partial}). However, this is true for a special form of these operators, and not expected to be true in general.} As should be clear from these examples, in most cases of realistic quantum systems, possibly with interaction terms, they are free up to a certain order (apart from the cases where they are constructed specifically to satisfy freeness, see next section), and if this number is sufficiently high, the exact DOS obtained from numerical calculations is quite close to that of the free probability prediction.   

\subsection{Generalised Anderson model with free random variables}\label{Anderson_free}
For the standard Anderson model with diagonal on-site disorder considered in the previous subsection, the disorder variables at different sites are assumed to be independent random variables, which,  however,  does not correspond to any specific relation between the component terms ($\BA$ and $\BB$) of the Hamiltonian.\footnote{In the previous subsection we did discuss the calculation of the DOS of the one-dimensional Anderson model with nearest-neighbour interaction by assuming that $\BA$ and $\BB$ are free; however as pointed out this assumption is valid only up to a certain order when the respective moments are matched with free probability prediction, i.e., the operators are partially free. The model we discuss in this section, following \cite{neu1995random,neu1995rigorous}, does have the property that $\BA$ and $\BB$ are asymptotically free. See also section \ref{sec_sum_AB} for a general discussion on the case where $\BA$ and $\BB$ are neither commuting nor asymptotically free and an effective procedure for obtaining the DOS by matching the fourth moment.} As a result, it is not possible to solve the Anderson model exactly. 
To circumvent this difficulty, a generalisation of the Anderson model was proposed in \cite{neu1995random,neu1995rigorous} (based on \cite{Wegner}), where at each lattice site $n$ different levels are assumed to be present \cite{Wegner}, and the resulting disordered variables at different sites are assumed to be free in the limit $n \rightarrow \infty$. These assumptions imply a relation between the operators $\BA$ and $\BB$ (in fact, one can show that they are asymptotically free), and it is then possible to solve the resulting model exactly.\footnote{For the use of the free random variables and the description of free cumulants based on non-crossing partitions to understand the spectra of Hamiltonians that have dynamical disorder, we refer the reader to \cite{neu1993self, neu1994spectra}.} 

For a $d$ dimensional lattice having $n$ electronic levels at each site $j$, the Hamiltonian under consideration is given by $H=\BA+\BB$,
where the part $\BB$ is a translation invariant and deterministic operator which is diagonal when written in terms of the basis $\ket{ \alpha}$ ($\alpha=1,2, \cdots n$), i.e., 
\begin{equation}
    \BB = \sum_{j, j^\prime, \alpha} J_{|j-j^\prime|} \ket{j \alpha}\bra{j^\prime \alpha}~,
\end{equation}
while the operator $\BA$ contains the on-site diagonal disorder given by 
\begin{equation}\label{WegnerA}
    \BA = \frac{1}{\sqrt{n}} \sum_{j, \alpha, \beta} h_j^{\alpha\beta} \ket{j \alpha}\bra{j \beta}~.
\end{equation}
In the original extension of the Anderson model with $n$ orbitals per site proposed by Wegner \cite{Wegner}, the matrices $h_j=\Big(\frac{h_j^{\alpha\beta}}{\sqrt{n}}\Big)_{\alpha,\beta=1}^{n}$ are Gaussian random matrices with independent entries for different sites $j \neq j^\prime$.  To understand the important generalisation of the Wegner model proposed in \cite{neu1995rigorous, neu1995random}, we diagonalise the random matrices $h_j$: $h_j=U_j \CH_j U_j^\dagger$, where $\CH_j$ is a diagonal matrix and $U_j$ is a unitary matrix, so that it can be seen that in the case of the Wegner model $\CH_j$ is a deterministic matrix having Wigner semicircle as the DOS in the limit $n \rightarrow \infty$, and $U_j$ is a unitary matrix drawn from the Haar measure on $U(n)$ (where the matrices for different lattice sites $U_j$ and $U_{j^\prime}$ are chosen independently from the Haar ensemble). Now replace the diagonal matrices $\CH_j$ by a single deterministic $n \times n$
Hermitian matrix (may not be diagonal) $\CH$, and consider the modified version of the operator $\BA$ \footnote{Compare this expression with  \eqref{WegnerA}.  As before, for different sites, the unitary matrices are independently drawn from the Haar measure on the unitary group.}
\begin{equation}
    \BA = \sum_{j, \alpha, \beta} (U_j \CH U_j^\dagger)_{\alpha\beta} ~\ket{j \alpha}\bra{j \beta}~, ~~ U_j \in U(n)~.
\end{equation}
Since the eigenbases of the matrices $h_j:=U_j \CH U_j^\dagger$ for different values of $j$ are randomly rotated against each other, the matrices $h_j$ are free for all different sites $j \neq 
j^\prime$ in the limit $n \rightarrow \infty$ with respect to the disorder average map $\varphi(\CO)=\avg{\CO}_d=n^{-1}\sum_\alpha \avg{\alpha|\CO|\alpha} $ (see the Example 1 in section \ref{sec_freeness}). 

Remarkably, the fact that $h_j$ are free for all different sites $j \neq j^\prime$,  \textit{directly implies that the component terms of the Hamiltonian $\BA$ and $\BB$ are also free with respect to the map}, 
$\varphi(.)=\avg{\avg{j|.|j}}_{d}$. This fact can be proved \cite{neu1995random} (see eqs. \eqref{free_def}, and \eqref{free_def2} for the definitions of freeness and asymptotic freeness, respectively) by first utilising the fact that, since $\BB$ is translationally invariant, for polynomials of $\BB$ with $\avg{j|q_k(\BB)|j}=0$, it also implies  $\avg{i(k)|q_{k}(\BB)|i(k)}=0$ for all $k$, and subsequently using the freeness of the $h_j$ matrices and $\avg{p_k(h_{i(k)})}_d=0$ (by assumption). 

Now, if we consider the ensemble-averaged single-particle Green function 
\begin{equation}
    \CG(j, j^\prime,z) = \bavg{\bavg {j\Big|\frac{1}{z-H}\Big|j^\prime}}_d = \bavg{\bavg{j\Big|\frac{1}{z-(\BA+\BB)}\Big|j^\prime}}_d~, 
\end{equation}
then the freeness of $\BA$ and $\BB$ implies that the diagonal part of 
this quantity, $G_{H}(z)=\CG(j, j,z) $ satisfies the following relations (see eqs. \eqref{subor_A} and \eqref{subor_B}) 
\begin{equation}\label{subor_anderson}
    G_H(z) = G_{\BB}\Big(z-R_{\BA}\big(G_{H}(z)\big)\Big) = G_{\BA}\Big(z-R_{\BB}\big(G_{H}(z)\big)\Big) ~,
\end{equation}
where $G_{\BA}$ is the diagonal part of the ensemble-averaged one-particle Green function of the operator $\BA$, and $R_{\BA}$ is the associated $R$-transform; similarly for $\BB$. One of these two relations can then be used to obtain the DOS of the Hamiltonian, given the DOS of the component operators (see \cite{neu1995random, neu1995rigorous} for several examples, including the case of Wegener's original model with Gaussian random matrices as $h_j$s).  

To summarise, in the previous three sections, we have discussed the use of methods of the free probability theory to  understand the DOS of quantum mechanical systems, focusing on three different but somewhat related models.   For the models discussed in sections \ref{ssec_LRMT} and \ref{Anderson_free}, the notion of asymptotic freeness is present due to the way these are constructed, and in a sense, using a similar `trick', namely, one takes a fixed reference operator and subsequently rotates independent copies of it by independent Haar unitaries - once per each Haar-Ising bond for the local random matrix Hamiltonian in section \ref{ssec_LRMT}, and once per each site for the generalised Anderson model considered in this subsection. Whereas the usual Anderson model with on-site or off-diagonal disorder discussed in section \ref{sec_dos_anderson} lacks this specific ingredient, and hence, its component operators fail to be asymptotically free.   

This concludes our discussion of examples where free additive convolution can be utilised to compute the DOS of quantum mechanical Hamiltonians. In the next section, we illustrate the  procedure of obtaining the density of states of a Hamiltonian of the form $\BC= \BA + \BB$
perturbatively using free probability by employing the assumption that the (finite-dimensional) component operators $\BA$ and $\BB$ are free from each other, and reanalyse the Anderson model discussed in this section to provide some analytical understanding of the computation of the DOS and the role played by the freeness of the component operators.

\section{A perturbative scheme for obtaining the density of states} 
\label{dos_perturb}

In this section, we develop a perturbative scheme which utilises the subordination formulas for the Cauchy transform to compute the DOS of a Hamiltonian of the form $H=\BA + \alpha \BB$  by assuming that the self-adjoint operators $\BA$  and $\BB$ are free from each other, and their respective DOS are known. The assumption of freeness between the two operators constituting the Hamiltonian, even though,  does not work in all situations, and is somewhat dependent on how one constructs $\BA$ and $\BB$ from a given Hamiltonian, can, in many cases, provide a remarkably accurate description of the DOS of complex quantum mechanical systems with disorder \cite{Chenprl, Welborn}. In the perturbative scheme that we are going to illustrate below, the parameter $\alpha$ acts as a small parameter with respect to which the perturbation series is developed, and in some cases, as we shall see, it provides a unique choice of the operators $\BA$  and $\BB$. Furthermore, the fact that one needs to know the analytical forms of the DOS of the constituent operators $\BA$ and $\BB$, also restricts the choices of these operators one can use realistically for a given $H$ to apply this approximation scheme. 


The procedure we elaborate below, starting from the subordination formulas in \eqref{subor_A} or \eqref{subor_B}, to set up a perturbation series for the $G_\BC(z)$, was used in \cite{Venturelli:2022hka} to calculate the DOS in the fractal phase of the Rosenzweig-Porter (RP) random matrix ensemble. 
It was subsequently employed in \cite{Jahnke:2025exd} to obtain analytical expressions for the first two orders of correction to a Gaussian DOS in the fractal phase of the RP model. 
In the following, first we keep the discussion general enough so that one can use it for a Hamiltonian that is a sum of two generic operators, then later use it to illustrate two important cases, (I) when the perturbation operator has a Wigner semicircle as the DOS, and (II) when the perturbation operator has an arcsine distribution as the DOS.

\subsection{A perturbation series for the Cauchy transform of $\BC$}

To see how to develop a perturbation series for the Cauchy transform of the operator $\BC$, we go back to the original problem of calculating the DOS of the operator $\BC=\BA + \alpha \BB$, where $\alpha$ is a small parameter.\footnote{Later, in our examples, we shall take the operator $\BC$ to be the Hamiltonian $H$ of some quantum system with disorder or belonging to a random matrix ensemble.} Since, $\alpha$ is assumed to be a small parameter, we expect the DOS of $\BC$ can be written in terms of the corrections to the DOS of the `initial' operator due to the `perturbation' operator $\BB$. On the other hand, if the operator $\BC$ is of the form, $\BC=\BA + \zeta \BB$, where the parameter $\zeta>1$, then one can rewrite it as $\BC=\zeta(\BB + \alpha \BA)$, where $\alpha=\zeta^{-1}$, now acts as the perturbation parameter, and the procedure described below can be used to compute the correction to the DOS of the `initial' operator $\BB$ due to the `perturbation' operator $\BA$.

To proceed further, it will be convenient to introduce the \textit{characteristic function} of the distribution $\rho_{\BA}(\lambda)$, defined, in our convention, as,
\begin{equation}\label{chiu}
\chi_\BA(u) = \int \text{d}\lambda ~\rho_{\BA}(\lambda)~ e^{- i \lambda u }\,. 
\end{equation}
Considering the inverse of this equation with an analogous equation to that of \eqref{dos_FT_U} for the operator $\BA$, we see that $\chi_\BA(u)$ is just the ensemble averaged (if the operator $\BA$ belongs to some ensemble of operators) trace of the evolution operator generated by the operator $\BA$; hence, $\chi_\BA(t) \equiv \mathcal{U}_{\BA}(t)$. 

To directly use the formula in \eqref{subor_A} in this case, we need the scaling property of the $R$-transform which says that for an operator $\mcO$, with a well-defined DOS, the corresponding $R$-transform follows the \textit{scaling law} $R_{\alpha \mcO}(z) = \alpha R_\mcO(\alpha z)$ \cite{Potters, livan2018introduction}. 
Now, using this scaling property of the $R$-transform along with the characteristic function defined above, we can rewrite eq. \eqref{subor_A} as follows,
\begin{align}\label{GC_exact}
	G_{\BC}(z) &=  \int \text{d}\lambda~ \frac{\rho_{\BA}(\lambda)}{z-\alpha R_{\BB} \big(\alpha G_{\BC}(z)\big)-\lambda} \nonumber \\
    &=i  \int_{0}^{\infty} ~ \text{d}u ~\chi_\BA(-u)  \exp \Big[
	-i u \Big(z- \alpha R_{\BB} \big(\alpha G_{\BC}(z)\big) \Big)\Big]~\\
    &=i  \int_{0}^{\infty} ~ \text{d}u ~\chi_\BA(-u) ~e^{-iuz}  \sum_{n=0}^{\infty} \frac{(iu \alpha)^n}{n!} \big[R_{\BB} \big(\alpha G_{\BC}(z)\big)\big]^n~.\label{GC_exa_2}
\end{align}
Since we want to separate the factors of $G_{\BC}(z)$ from both sides, we use the fact that the $R$-transform obtained from the DOS  of an operator is the generator of its free cumulants (see the expansion in eq. \eqref{free_cumuluants}).  
Substituting eq. \eqref{free_cumuluants} in the last expression for $G_{\BC}(z)$, we have, the following general equation for it,  
\begin{align}\label{GC_kappa}
	G_{\BC}(z) &= \sum_n \frac{i^{n+1}\alpha^n}{n!} \Big(\sum_{j=1}^{\infty}\kappa_j(\BB)\big(\alpha G_{\BC}(z)\big)^{j-1}\Big)^n  \int_{0}^{\infty} ~ \text{d}u ~\chi_\BA(-u) ~ u^n ~e^{-iuz}\\
    &=\sum_n \frac{(-1)^n \alpha^n}{n!} \Big(\sum_{j=1}^{\infty}\kappa_j(\BB)\big(\alpha G_{\BC}(z)\big)^{j-1}\Big)^n \partial_z^n G_{\BA}(z) \,.
\end{align}
In the final expression above, we have used the following relation between the characteristic function of the distribution $\rho_{\BA}(\lambda)$ and its Cauchy transform (see the expression in \eqref{G_evolution}) 
\begin{equation}
    G_\BA(z)= i \int_0^{\infty} \text{d}u ~ \chi_\BA (-u)\, e^{-i u z}~,~~\Im z<0~.
\end{equation}

The final expression in \eqref{GC_kappa} is a self-consistency equation for  $G_{\BC}(z)$, written as a series expansion. Our strategy in the following would be to treat the parameter $\alpha$ as a small parameter in a perturbation expansion of this quantity, and try to solve for it. 

\subsection{Case 1: The perturbation operator $\BB$ has the Wigner semicircle density of states.}\label{W_perturb}
For a generic self-adjoint operator $\BB$ with a given DOS, the corresponding $R$-transform can be quite complicated; therefore, applying the exact expression in \eqref{GC_kappa} in a generic situation may not be very helpful. Let us first consider the simplest scenario, a particular expression for the DOS of $\BB$ so that the free cumulant expansion of $R_{\BB}(z)$ has only one term, i.e., $R_{\BB}(z)=\kappa_2 z$ (where $\kappa_2>0$, and we scale the distribution in such a way that $\kappa_2=1$). In that case, the DOS of $\BB$ is the Wigner semicircle of the form $\rho_{\BB}(\lambda) = \frac{1}{2 \pi} \sqrt{4-\lambda^2}$.  This is the analogue of the Gaussian distribution for the case of classical probability to the non-commutative operators.

For this case, eq. \eqref{GC_kappa} simplifies to 
\begin{align}\label{GC_sum}
  G_{\BC}(z) &=  \sum_n\frac{i^{n+1}}{n!} \alpha^{2n}\,\big(G_{\BC}(z)\big)^n \int_{0}^{\infty}  \text{d}u ~u^n \chi_\BA(-u) e^{-iuz} \,,\nonumber \\
  &=\sum_n \frac{(-1)^n}{n!} \alpha^{2n}\, \big(G_{\BC}(z)\big)^n\, \partial_z^n G_{\BA}(z)\,.
\end{align}

The next step is to solve for $G_{\BC}(z)$  from the series expansion above for different orders of the small parameter $\alpha$. At $\mathcal{O}(\alpha^0)$, we have the DOS of $\BC$ same as that of $\BA$, since, 
\begin{equation}
	G_{\BC}^{(0)}(z) = i  \int_{0}^{\infty} ~ \text{d}u ~\chi_\BA(-u) ~e^{-iuz}~=G_{\BA}(z)\,.
\end{equation}
The first order correction to the Cauchy transform $G_{\BC}(z)$ of $\rho_\BC (\lambda)$ is given by
\begin{align}\label{GC_1}
	G_{\BC}(z) \approx G_{\BC}^{(1)}(z) + \mathcal{O}(\alpha^4)~,~~\text{where}~~ ~G_{\BC}^{(1)}(z) ~&= G_{\BC}^{(0)}(z) - \alpha^2  G_{\BC}^{(0)}(z) ~ \partial_z G_{\BC}^{(0)}(z) \\&= G_{\BC}^{(0)}(z) - \frac{\alpha^2}{2}\partial_z \big(G_{\BC}^{(0)}(z) \big)^2 ~,
\end{align}
while the Cauchy transform with the 2nd order correction\footnote{Note that the term we call first-order correction here is proportional to $\alpha^2$. Similarly, the second-order correction terms are proportional to at least the fourth power of $\alpha$, and so on. One can, of course, expand these individual terms and obtain a power series in $\alpha$, and since $\alpha<1$, we expect the higher order terms to have progressively smaller effects on the DOS.} is 
given by \cite{Jahnke:2025exd}
\begin{equation}
    G_{\BC}(z) \approx G_{\BC}^{(2)}(z) + \mathcal{O}(\alpha^6)~,
\end{equation}
with 
\begin{align}\label{GC_2}
G_{\BC}^{(2)}(z)= G_{\BC}^{(0)}(z) - \alpha^2  G_{\BC}^{(0)}(z) \partial_z G_{\BC}^{(0)}(z)
+ \frac{\alpha^4}{2} \Big(G_{\BC}^{(0)}(z) - \alpha^2  G_{\BC}^{(0)}(z) \partial_z G_{\BC}^{(0)}(z)\Big)^2 \partial_{zz} G_{\BC}^{(0)}(z)\,.
\end{align}
Continuing in a similar manner, all the higher-order correction terms can be obtained from the Cauchy transform of the operator $\BA$ and its derivatives. But for our purposes in this section, the above expression for the Cauchy transform with second-order correction will be sufficient. 
Finally, the approximate DOS at each order for $C$ can be computed from $G_{C}(z)$ using the Stieltjes inversion formula in \eqref{dos_from_G}.

Before proceeding further, we note an important point regarding the use of the formulas in eqs. \eqref{subor_A} or \eqref{subor_B} to compute the DOS of the operator $C$. The form of the operator 
$\BC=\BA + \alpha \BB$, with $\alpha$ being a small parameter, makes eq.  \eqref{subor_A} a natural choice. As we have discussed, since it is difficult to obtain the exact expression of $ G_{\BC}(z)$ (and hence the corresponding DOS, $\rho_{\BC}(\lambda)$), by performing the sum, e.g., in \eqref{GC_sum}, the approximate expressions of different orders in eq. \eqref{GC_1} and \eqref{GC_2} essentially give us different orders (in the small parameter $\alpha$) of corrections to the DOS of the operator $\BA$ due to the operator $\BB$. On the other hand, it should be evident that if the operator (or the Hamiltonian) under consideration is of the form, $\BC= \alpha \BA + \BB$, one has to use the formula in \eqref{subor_B},  which would give corrections to the DOS of the operator $\BB$ due to the operator $\BA$. We should also note that, in a given model, the approximate DOS obtained using the perturbation series may not work in the entire range of its parameter space. E.g., as we shall see below, for a Hamiltonian belonging to the RP random matrix ensemble, this method gives a quite good approximation to the DOS in the fractal phase \cite{Venturelli:2022hka, Jahnke:2025exd}, where it is essentially a deformed Gaussian distribution; however, in the ergodic phase, the above method does not work, since the DOS, in this phase, is a deformed Wigner semicircle, hence one needs to consider correction to the DOS of the operator $\BB$, not that of $\BA$.   

\textbf{Example 1: Approximate DOS of the RP model.} Let us now apply the eqs. \eqref{GC_1} and \eqref{GC_2} to compute the DOS of the RP model. A Hamiltonian belonging to the RP random matrix ensemble can be written as $H=\BA+\alpha \BB$, where the parameter $\alpha=N^{-\gamma/2}$ \cite{rosenzweig1960repulsion, kravtsov2015random}.  The elements of the diagonal matrix $\BA$ are drawn from a Gaussian distribution with zero mean and variance $4/N$, and $\BB$ is an $N \times N$ random matrix drawn from one of the canonical Gaussian random matrix ensembles, which we chose here to be the Gaussian orthogonal ensemble with variance $2/N$. The parameter $\gamma>0$ determines different phases of this ensemble: $0 \leq \gamma \leq 1$ corresponds to the ergodic phase, $1 < \gamma < 2$ is the non-ergodic extended (fractal) phase, while $\gamma>2$ corresponds to the localised phase.

The DOS of $\BA$ is given by $\rho_\BA(\lambda) = \frac{1}{\sqrt{2 \pi  \sigma^2}}  e^{-\frac{\lambda^2}{2 \sigma^2}}$, hence its characteristic function is given by $\chi_\BA(u) = e^{-\frac{1}{2}u^2 \sigma^2}$. This can be used to obtain the Cauchy transform of the operator $\BA$ in terms of the Dawson function $D(x) = e^{-x^2} \int_{0}^{x}e^{t^2}~\text{d}t$ as (with our convention for $G(z)$)
\begin{equation}\label{resolv_Gaus}
	G_{\BC}^{(0)}(z) = G_{\BA}(z) = \sqrt{ \frac{\pi}{2 \sigma^2}} \Bigg[-i e^{-\frac{z^2}{2 \sigma^2}} +  \frac{2}{\sqrt{\pi}} D \bigg(\frac{z}{\sqrt{2} \sigma}\bigg)\Bigg]~.
\end{equation}
One can use this expression for $G_{\BC}^{(0)}(z)$ in eqs. \eqref{GC_1} and \eqref{GC_2} to obtain the first and second-order corrections to $G_{\BC}(z) $, and subsequently use them to obtain the DOS of the RP model for $\gamma>1$. The resulting DOS can be checked to provide a very good approximation to the numerical result. We refer the reader to \cite{Jahnke:2025exd, Venturelli:2022hka} for further details. 

This procedure can also be used to compute the approximate DOS of a generalised version of the standard RP model with Hamiltonian of the form \cite{Gill:2025upp} $H=\BA + \alpha \BB_\beta$, where, as before, $\BA$ is a diagonal matrix with elements drawn from independent Gaussian distributions with zero mean and fixed variance. However, now the matrix $\BB_\beta$, instead of being drawn from one of the three classical Gaussian ensembles, is drawn from the tridiagonal Gaussian $\beta$-ensemble  \cite{dumitriu2002matrix, forrester2010log} with a generic value of Dyson index $\beta>0$ and therefore does not have rotational invariance of classical Gaussian random matrix ensembles. We refer the reader to \cite{Gill:2025upp} for further details of the numerical results for the DOS, and its comparison with the free probability predictions in this model. 

\textbf{Example 2: Approximate density of states of the Anderson model with high off-diagonal interaction strength} 

We can similarly apply this approximation scheme to obtain the approximate DOS of the Anderson model in \eqref{H_anderson} for large values of the off-diagonal interaction strength $J$. As far as we are aware, perturbation series for the Cauchy transform based on the subordination formulas obtained from free convolution have not been used previously in the literature to compute the DOS of the Anderson Hamiltonian. Therefore, to the best of our knowledge, the results reported in the rest of this section are new. As we shall see, this approach provides us with some analytical understanding of the DOS of this Hamiltonian.\footnote{Also note that, even though we do not discuss that case in the following, one can use the subordination formulas in \eqref{subor_anderson} for the Anderson model with free random variables to construct a perturbative expression for (the diagonal part of) $G_{H}(z)$, and hence the compute the approximate DOS from it.} 

For convenience, here we obtain the DOS of a scaled version of the Hamiltonian in \eqref{H_anderson}, which in our notation is given by $H=\BA+\alpha \BB$, where $\alpha=1/J$.\footnote{ The DOS of the unscaled Hamiltonian in \eqref{H_anderson} can be obtained by utilising the scaling relation for the Cauchy transform $G_{\alpha \mcO}(z)=\frac{1}{\alpha}G_{\mcO}(z/\alpha)$.} Note that, this choice of the operators $\BA$ and $\BB$ corresponds to the scheme I in \cite{Chenprl}. From the results reported in this reference, we expect that the free probability would provide a good approximation to the DOS of the Anderson model. 

The elements of the diagonal matrix $\BB$ are assumed to be drawn from the distribution $\rho_{\BB}(\lambda)$, which we take here to be the Wigner semicircle\footnote{In the usual version of the Anderson model, this is taken to be a Gaussian distribution. Note that even though the DOS of this operator is a Wigner semicircle, the eigenvalues are uncorrelated, unlike the case of random matrices having the same DOS. This can be thought of as an example of the so-called decorrelated ensembles \cite{Camargo:2025zxr}. } of the form $\rho_{\BB}(\lambda)=\frac{1}{2\pi}\sqrt{4-\lambda^2}$. 
The matrix $\BA$ has a tridiagonal form with all the non-zero elements being equal to unity. In the limit $N \rightarrow \infty$, the DOS of this matrix converges to the arcsine distribution given in \eqref{arcsine} \cite{Chenprl, strang1999discrete}. The Cauchy transforms of these two distributions can be easily computed and  are given by, respectively,
\begin{equation}\label{Cau_tr_ss_arc}
    G_{\BB}(z)=\frac{1}{2}\Big(z-\sqrt{z^2-4}\Big)~,~~\text{and}~~G_{\BA}(z)=\frac{1}{\sqrt{z^2-4}}~.
\end{equation}
Since these two Cauchy transforms are related by the simple relation, $(G_{\BA}(z))^3=\frac{1}{2}G_{\BB}^{\prime \prime}(z)$ (where the prime denotes a derivative with respect to $z$), the general formula for the Cauchy transform of the Hamiltonian in  \eqref{GC_sum} can be written in terms of $G_\BB(z)$ as 
\begin{equation}
    G_{H}(z) = \sum_n \frac{(-1)^n}{n!2^{1/3}} \alpha^{2n}\, \big(G_{H}(z)\big)^n\, \partial_z^n \big(G_\BB^{\prime \prime}(z)\big)^{1/3}~.
\end{equation}
In particular, at zeroth order we have, $G_{H}^{(0)}(z) = G_{\BA}(z)=(\frac{1}{2}G_{\BB}^{\prime \prime}(z))^{1/3}$.

We can now obtain the expressions for the corrections to the Cauchy transform of the arcsine DOS of the matrix $\BB$ due to $\BA$. From eqs. \eqref{GC_1} and \eqref{GC_2}, the Cauchy transforms with first and second order corrections are given by, respectively, 
\begin{equation}\label{GH_1_diso}
    ~~G_{H}^{(1)}(z) = \frac{1}{\sqrt{z^2-4}} \Bigg(1+\frac{\alpha^2 z}{(z^2-4)^{3/2}}\Bigg)~,
\end{equation}
and 
\begin{align}\label{GH_2_diso}
       G_{H}^{(2)}(z) &= G_{H}^{(1)}(z) +\frac{\alpha^4}{2} \big(G_{H}^{(1)}(z)\big)^2 \frac{2(2+z^2)}{(z^2-4)^{5/2}}~\\
       &=\frac{1}{\sqrt{z^2-4}} \Bigg(1+\frac{\alpha^2 z}{(z^2-4)^{3/2}}\Bigg)+ \alpha^4 \frac{z^2+2}{(z^2-4)^{15/2}}\Big((z^2-4)^2+\alpha^2 z \sqrt{z^2-4}\Big)^2~.
\end{align}
The first and second order corrections to the DOS can be obtained from these expressions for the Cauchy transform using the Stieltjes inversion formula, eq. \eqref{dos_from_G}. It can be easily checked that the contribution of the correction term (the second term in \eqref{GH_1_diso}) in $G_{H}^{(1)}(z) $ in the DOS actually vanishes, i.e., in the first order, the DOS is exactly the same as the arcsine distribution. 

\begin{figure}[h!]
		\centering
          \begin{tabular}{cc}
		\includegraphics[width=0.5\textwidth]{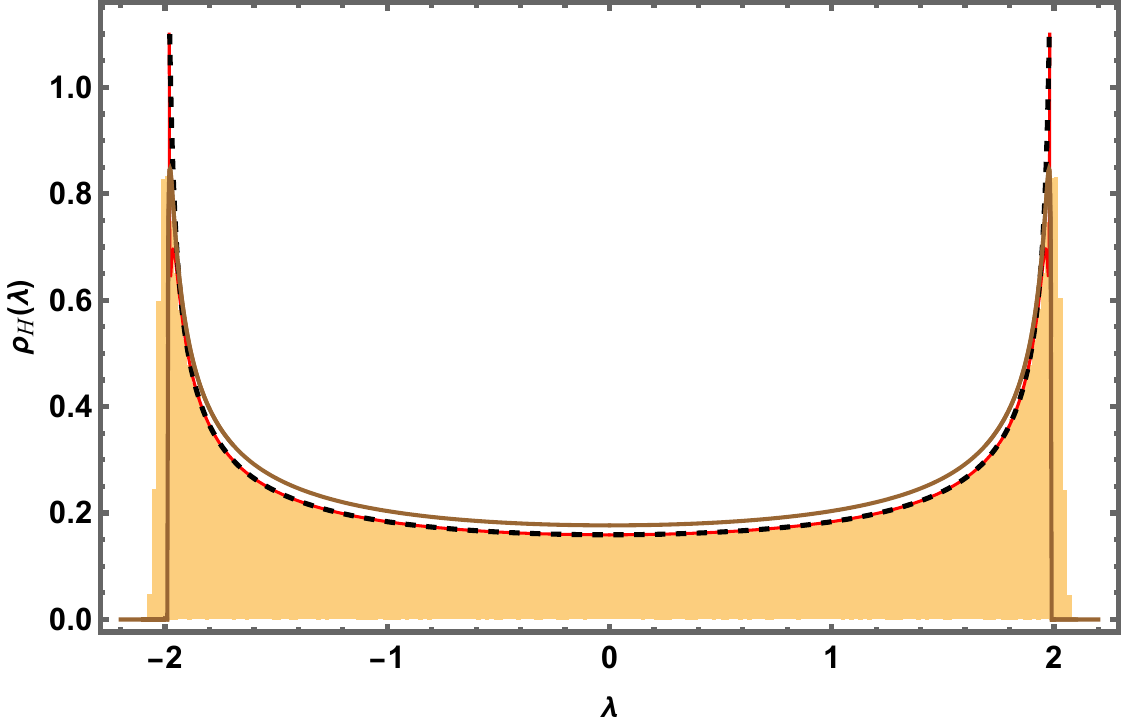}&
            \includegraphics[width=0.5\textwidth]{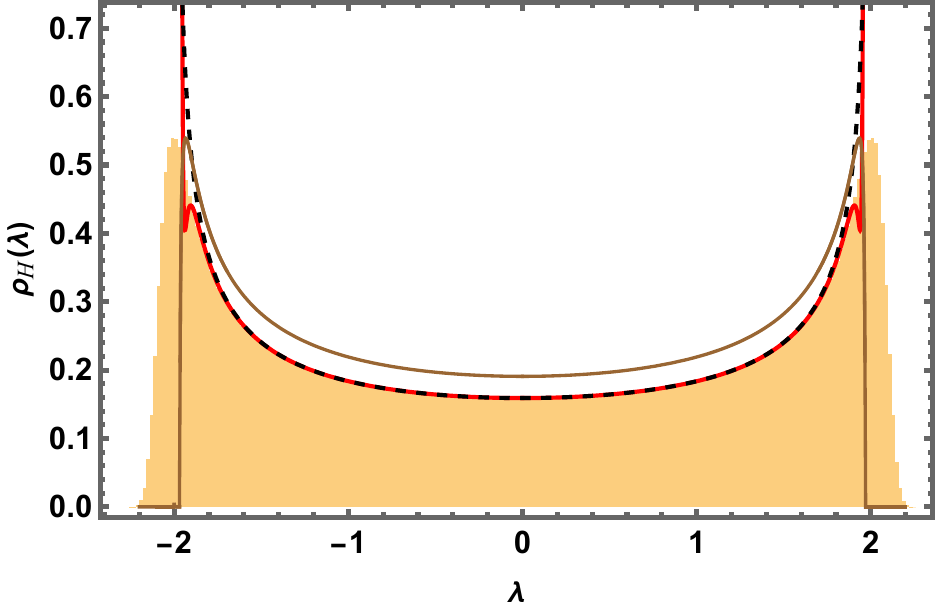}
        \end{tabular}
		\caption{DOS of the Anderson model with high off-diagonal interaction strength and its comparison with the free probability approximation. Here we have set $J=10$ (left plot) or $J=5$  (right plot), and the size of the Hamiltonian is $N=2000$. The numerical results show an average over 1000 independent realisations of the Hamiltonian. The red curve indicates $\rho_{H}(\lambda)$ obtained from the second-order Cauchy transform in \eqref{GH_2_diso}, while the black dashed curve represents the arcsine distribution. For comparison, with the brown curves in both the plots, we have also shown a free compressed version of the arcsine distribution (with the associated Cauchy transform in \eqref{Cauchy_KM_com}, $\eta=2$) with free compression parameter, for the left plot $0.895$, and $0.82$ for the right plot.}
		\label{fig:dos_J_high}
	\end{figure}

In Fig. \ref{fig:dos_J_high} we have shown a comparison of the numerical DOS along with the one obtained from this approximation scheme for different values of $J$, i.e., the parameter $\alpha$. For sufficiently large values of $J$, i.e., small $\alpha\ll 1$, the DOS is almost identical to the initial arcsine distribution of the operator $\BA$ (see the left panel of Fig. \ref{fig:dos_J_high}). However, as the value of $J$ is decreased, the edges of the DOS of $H$ cross the edges $\pm 2$ of the initial arcsine distribution, and go to zero more smoothly. Since all orders of corrections to the Cauchy transform $G_{H}(z)$ have a sharp singularity at $\pm 2$, the approximation scheme described above can not capture the smooth behaviour of $\rho_{H}(\lambda)$ towards the edges of the spectrum. 

It is also interesting to make a comparison between the numerical DOS and a free compressed version of the arcsine distribution (see Section \ref{free_compression}), since in the latter case as well, for a suitably chosen compression parameter range, the distribution function does not diverge at the edges. In Fig. \ref{fig:dos_J_high}, by the brown curve, we have shown the free compressed arcsine distribution. As can be seen, by suitably choosing the value of the compression parameter (see Cauchy transform in \eqref{Cauchy_KM_com}, and the corresponding compressed distribution \eqref{comr_KM} with $\eta=2$), the distribution obtained via free compression of the arcsine distribution does provide a reasonable approximation to the numerical DOS, specifically for sufficiently large values of the off-diagonal interaction strength.

\subsection{Case 2: The perturbation operator $\BB$ has arcsine density of states.}

So far, we have considered the relatively simple case where the perturbation operator $\BB$ has a semicircle DOS, so that the corresponding $R$-transform is just $z$. In this section, our goal is to extend the approximation scheme developed above for other cases where the analytical expression of the $R$-transform is known and has more than one non-zero free cumulant. As we shall see, this will be useful to approximate the DOS in the bulk of the spectrum of the Anderson Hamiltonian \eqref{H_anderson} with low off-diagonal interaction strength $J\ll 1$. Keeping this example in mind, here we assume that the operator $\BB$ has the arcsine distribution as the DOS. Furthermore, we shall show that for this choice of the perturbation $\BB$, the lowest order (in the perturbation parameter, here $\alpha=J$)  correction to the Cauchy transform is the same as when the perturbation operator has Wigner semicircle DOS with a rescaled parameter, irrespective of the initial operator $\BA$. 

To obtain the expression for the Cauchy transform of $\BC=\BA+\alpha \BB$, we start from the expression in \eqref{GC_exact}, which we rewrite as
\begin{align}\label{GC_kappa_2}
	G_{\BC}(z) 
    &=\sum_n \frac{(-1)^n \alpha^n}{n!} \big[R_{\BB} \big(\alpha G_{\BC}(z)\big)\big]^n~ \partial_z^n G_{\BA}(z) \,.
\end{align}
Now for the arcsine distribution, the $R$-transform can be obtained to be \cite{nica2006lectures, mingo2017free}
\begin{equation}
    R_{\BB}(z)=\frac{1}{z} \Big(-1+\sqrt{1+4 z^2}\Big)~. 
\end{equation}
Substituting this expression in \eqref{GC_kappa_2}, we have 
\begin{align}\label{GC_kappa_3}
	G_{\BC}(z) 
    &=\sum_n \frac{(-1)^n }{n!} ~ \Big(\partial_z^n G_{\BA}(z)\Big)~\Big[\frac{1}{G_{\BC}(z)}\Big(-1+\sqrt{1+4 \alpha^2G_{\BC}^2(z) }\Big)\Big]^n \,.
\end{align}
Therefore, the first correction (coming from the $n=1$ term in the last expression) to the DOS of $\BA$ due to the operator $\BB$ is given by (as before, the zeroth order expression is the same as that of the operator $\BA$, i.e., $G_{\BC}^{(0)}(z)=G_{\BA}(z)$)
\begin{align}\label{GC_approx}
    G_{\BC}(z) & \approx G_{\BC}^{(0)}(z) - \frac{1}{ G_{\BC}^{(0)}(z) }\Big(-1+\sqrt{1+4 \alpha^2 (G_{\BC}^{(0)}(z)) ^2 }\Big) ~ \partial_z G_{\BA}(z)~.
\end{align}
Note that, unlike the case 1 considered in section \ref{W_perturb}, here we do not write the corrections as a power series in different orders of $\alpha$. In principle, one can expand the expression in \eqref{GC_kappa_3} in a power series in $\alpha$, and obtain corrections in different orders of $\alpha$. E.g., the expression in 
\eqref{GC_kappa_3} can be rewritten formally as \footnote{This expression is essentially the same expression as one would get by expanding $R_{\BB}(z)$ here in terms of its free cumulants. In this case, there are an infinite number of them.}
\begin{align}\label{GC_kappa_4}
	G_{\BC}(z) 
    &=\sum_n \frac{(-1)^n }{n! G_{\BC}^n(z)} ~ \partial_z^n G_{\BA}(z)~\Big[\sum_{k=1}^\infty \frac{(1/2)_{k}}{k!} (4\alpha^2)^k G_{\BC}^{2k}(z)\Big]^n \,,
\end{align}
where $(r)_k$ denotes the Pochhammer symbol. Therefore, the lowest order in $\alpha$ contribution coming from the $n=1$ term (as in \eqref{GC_approx}) is of $\CO(\alpha^2)$, and $n=2$ or other higher order $n$ do not have any contribution of $\CO(\alpha^2)$ or lower order.  For our purposes here, the expression in \eqref{GC_approx} will be sufficient in most cases; however, it is interesting to compare the nature of the correction terms in this case with the one considered previously in Case 1, i.e., when the perturbation operator $\BB$ has the Wigner semicircle DOS. Expanding the expression in \eqref{GC_approx} in powers on $\alpha$, keeping upto $\CO(\alpha^2)$ terms, we get,  
\begin{align}\label{GC_1st_arc}
    G_{\BC}(z) & \approx G_{\BC}^{(0)}(z) - 2 \alpha^2 G_{\BC}^{(0)}(z)~ \partial_z G_{\BC}^{(0)}(z) = G_{\BC}^{(0)}(z) - \alpha^2 \partial_z \big(G_{\BC}^{(0)}(z)\big)^2~.
\end{align}
Comparing this with the expression in \eqref{GC_1}, we see that, apart from a factor of two, the two expressions are almost identical.  This indicates that, for a given operator $\BA$, if a perturbation operator $\BB$ (which is scaled by a factor of $\alpha < 1$) is added to it, then the first-order correction to the Cauchy transform is similar when $\BB$ has either a semicircle or an arcsine DOS. If $\alpha \ll 1$, then one can, in fact, absorb the extra factor of $2$ in \eqref{GC_1st_arc} in this parameter, so that the correction terms with the modified perturbation parameter would become identical with the expression in \eqref{GC_1}. This relation can be traced to be associated with the expressions for the $R$-transform of the DOS in the two cases. 

\textbf{Example 3: Approximate density of states of the Anderson model with low off-diagonal interaction.}

Returning to the Anderson Hamiltonian, we now have $\alpha=J$, where the diagonal matrix $\BA$ has elements drawn from the Wigner semicircle, and $\BB$ is a tridiagonal matrix with all diagonal elements zero, and all sub-diagonal elements unity. In that case, the expression for the approximate $G_{H}(z)$ we obtain from eq. \eqref{GC_approx} is given by, 
\begin{equation}\label{GH_approx_2}
    G_{H}(z) \approx \frac{1}{2\sqrt{z^2-4}} \Big(2-z^2+z\sqrt{z^2-4}+ 2\sqrt{1+\alpha^2 \big(z-\sqrt{z^2-4}\big)^2}\Big)~. 
\end{equation}

	\begin{figure}[h!]
		\centering
          \begin{tabular}{cc}
		\includegraphics[width=0.5\textwidth]{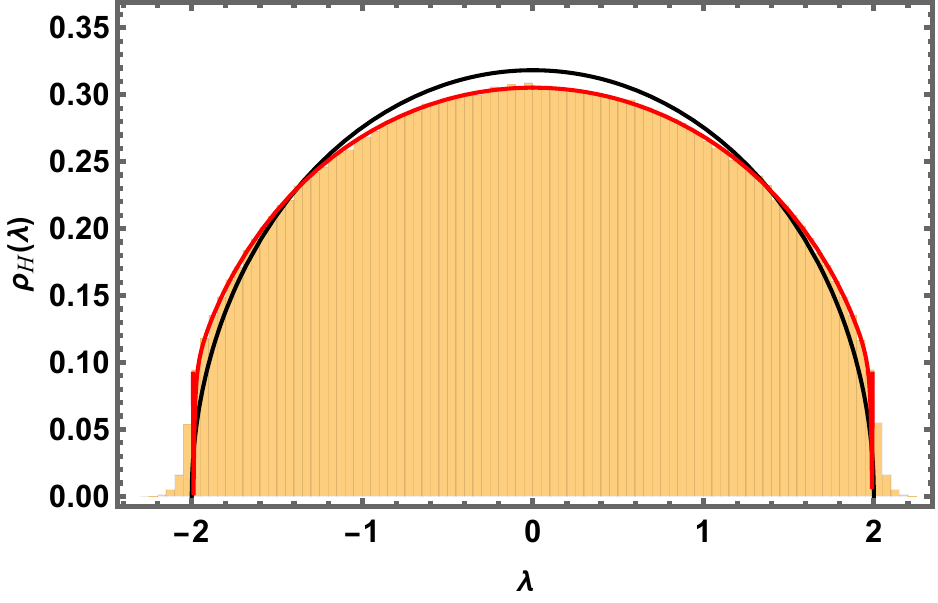}&
            \includegraphics[width=0.5\textwidth]{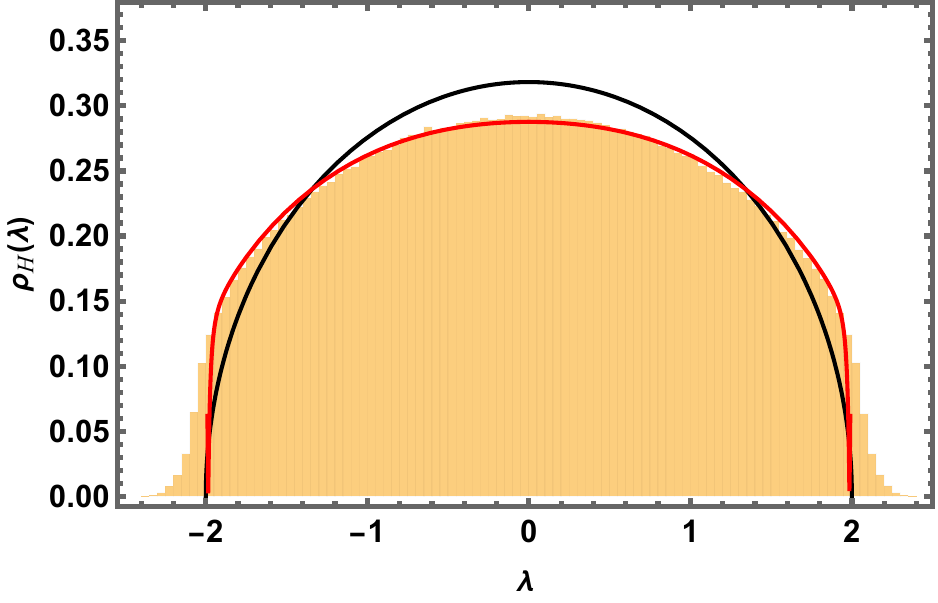}
        \end{tabular}
		\caption{DOS of the Anderson model with low off-diagonal interaction strength and its comparison with the approximate DOS obtained using the subordination formula. Here we have set $J=0.2$ (left plot) or $J=0.3$  (right plot), and the size of the Hamiltonian is $N=2000$. The numerical results show an average over 1000 independent realisations of the Hamiltonian. The red curve indicates $\rho_{H}(\lambda)$ obtained from the second-order Cauchy transform in \eqref{GH_2_diso}, while the black curve represents the Wigner semicircle distribution. The approximate formula provides a good description of the exact DOS in the bulk of the spectrum; however, it fails near the edges (we have not shown the sharp edges in the red curves for clarity).}
		\label{fig:dos_J_low}
	\end{figure}

In Fig. \ref{fig:dos_J_low}, we have shown a comparison of the numerically obtained DOS of the Anderson model with low off-diagonal interaction strength with the one obtained from the Cauchy transform in \eqref{GH_approx_2}. As can be seen, the approximate DOS provide a good description of the exact DOS in the bulk of the spectrum; however, near the edges of the spectrum, it fails due to the fact that the Wigner semicircle has edges at $\lambda=\pm 2$, which is manifested as a singularity in $G_{H}(z)$. We also mention that this is a problem of the analytical approximation we have used, and possibly not due to our assumption that the operators $\BA$ and $\BB$ are free, since numerically obtained DOS from free convolution provides a very good description of the exact DOS, even near the edges \cite{Chenprl, Welborn}. 

\textbf{Example 4: Approximate density of states of the Anderson model with Gaussian disorder and low off-diagonal interaction.}
As we have discussed above, when the diagonal elements of the Anderson Hamiltonian \eqref{H_anderson} are drawn from the Wigner semicircle, even though the analytical expression for the Cauchy transform $G_{H}(z)$ is simpler, it can not describe the numerical DOS well towards the edges of the spectrum due to the sharp edges of the semicircle.  To see that in the case the operator $\BA$ has smooth edges, the approximation scheme results in a smooth DOS, we now consider the case where the diagonal elements are drawn from a Gaussian distribution with mean zero and variance unity.  The expression for the Cauchy transform of this distribution is given in eq. \eqref{resolv_Gaus}, which is also the zeroth-order Cauchy transform of the Anderson Hamiltonian here. Using this in \eqref{GC_kappa_3}, we get the corrected Cauchy transforms of the Hamiltonian. 

	\begin{figure}[h!]
		\centering
          \begin{tabular}{cc}
		\includegraphics[width=0.5\textwidth]{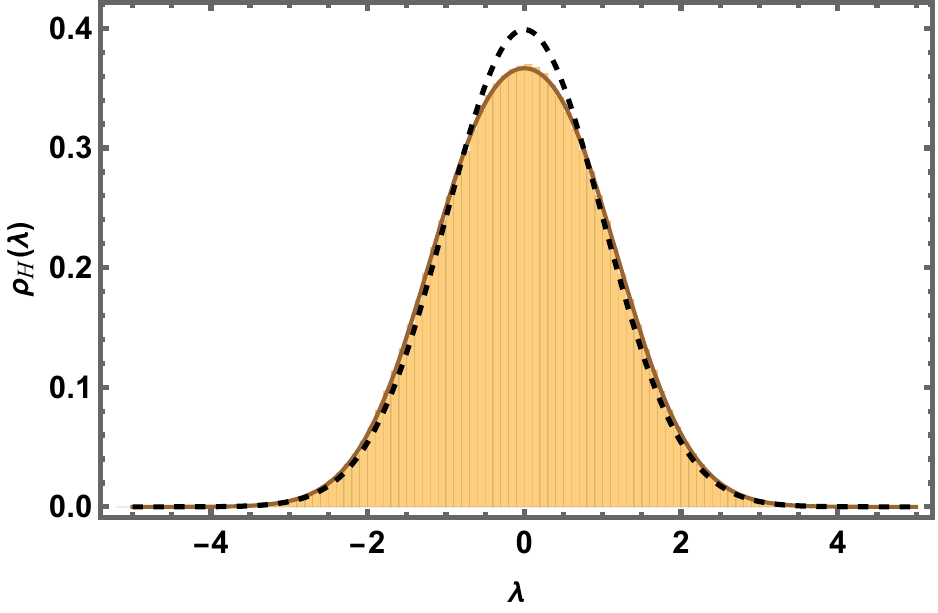}&
        \includegraphics[width=0.5\textwidth]{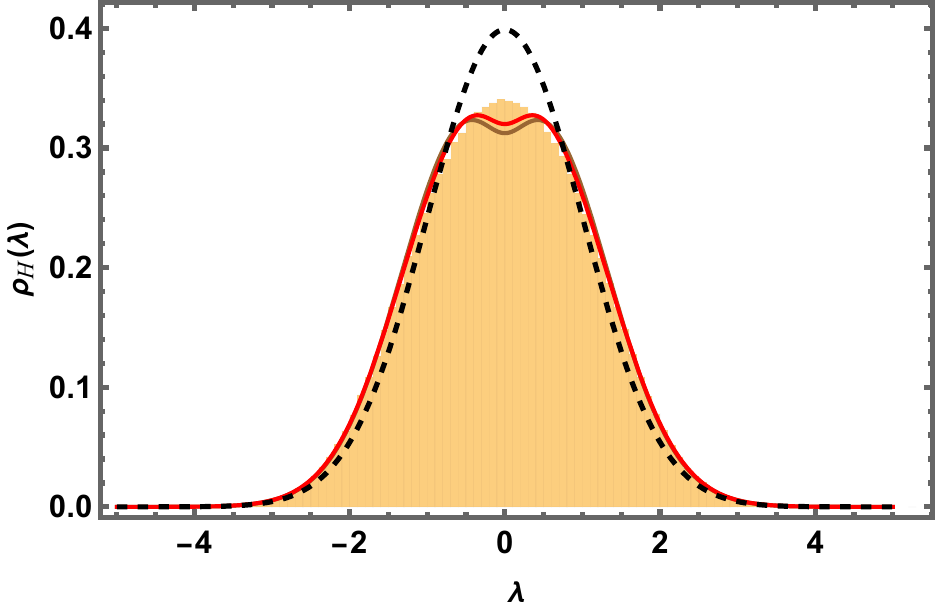}
        \end{tabular}
		\caption{DOS of the Anderson model with Gaussian on-site disorder ($\sigma=1$) and low off-diagonal interaction strength, and its comparison with the approximate DOS obtained using the subordination formula. Here we have set $J=0.2$ (left plot) or $J=0.3$  (right plot), and the size of the Hamiltonian is $N=2000$. The numerical results show an average over 1000 independent realisations of the Hamiltonian. The brown curve indicates $\rho_{H}(\lambda)$ obtained from the Cauchy transform in \eqref{GC_approx}, while the black dashed curve represents the Gaussian distribution. In the right panel, the red curve represents the DOS obtained from \eqref{GC_kappa_3} by retaining up to $n=2$ terms.}
		\label{fig:dos_J_gaussian_low}
	\end{figure}

In Fig. \ref{fig:dos_J_gaussian_low}, we have shown the numerical DOS along with the approximate DOS for the Hamiltonian in \eqref{H_anderson} whose diagonal elements are drawn from a Gaussian distribution with mean zero and variance unity. For a relatively small value of $J \ll 1$ (e.g., see the left plot with $J=0.2$), the first-order expression in \eqref{GC_approx} provides a very good approximation to the exact DOS.  However, as $J$ is increased, the approximation begins to deteriorate in the middle of the spectrum.  In that case, it is possible to provide a more accurate approximation by considering higher $n$ terms in \eqref{GC_kappa_3}. In the right panel of Fig. \ref{fig:dos_J_gaussian_low}, we show the approximate DOS by taking up to $n=2$ terms (the orange curve) for $J=0.3$, which, as can be seen, provides a slightly better approximation than the one obtained by considering only $n=1$ in the middle portion of the spectrum.

\section{Density of states  of sum of two generic $N \times N$ matrices}\label{sec_sum_AB}

So far in using the free convolution to calculate the DOS of an operator, which is a sum of two individual operators, we have assumed that the  $\BA$ and $\BB$ are asymptotically free from each other. However, freeness is a very strict condition to impose on these operators, which roughly demands that the eigenvectors of $\BA$ and $\BB$ are related by a $\beta$-Haar orthogonal transform.  In general, specifically, when the operator $\BC$ under consideration is Hamiltonian of some quantum system, the $N \times N$ Hermitian matrices, $\BA$ and $\BB$ may not be completely free from each other,\footnote{We have seen an example of this for the case of the Anderson model with nearest neighbour interaction and on-site diagonal disorder. Only for the case of the generalised Anderson model with free random variables considered in section \ref{Anderson_free}, $\BA$ and $\BB$ are asymptotically free.} a situation sometimes referred to as the \textit{partial freeness} \cite{chen2012partial, movassagh2010isotropic}. In this general case, when the operators $\BA$ and $\BB$ are neither commutative nor asymptotically free, an approach to obtain the DOS from the knowledge of the moments has been developed in \cite{movassagh2010isotropic, movassagh2017eigenvalue} by matching the fourth moment of the Hamiltonian between different situations. Here we briefly review this approach. 

These matrices can be diagonalised by suitable symmetry transforms, i.e., $\BA=\CQ_a^{-1}\Lambda_a\CQ_a$ and $\BB=\CQ_b^{-1}\Lambda_b\CQ_b$, where $\Lambda_a$ and $\Lambda_b$ are two diagonal matrices containing the eigenvalues of the two matrices, and $\CQ_i$ are $\beta$-orthogonal matrices which diagonalise $\BA$ and $\BB$. The Dyson index $\beta$ can take values $1,2,4$ depending on whether the symmetry class is orthogonal, unitary or symplectic, respectively. We are interested in the DOS of the sum $\BC=\BA+\BB$, which can be rewritten as,  $\BC=\Lambda_a+\CQ^{-1}\Lambda_b \CQ$, where $\CQ=\CQ_b\CQ_a^{-1}$. The classical and free additive convolutions are two extremes of this general $\CQ$. Namely, for the classical convolution, the matrices are assumed to commute with each other, and the transform matrix $\CQ$ in that case is just a random permutation matrix $(\Pi)$, hence, $\BC_c=\Lambda_a+\Pi^{-1}\Lambda_b \Pi$. On the other hand, when the matrices $\BA$ and $\BB$ are asymptotically free, $\CQ$ is a $\beta$-Haar orthogonal matrix, i.e., $\BC_f=\Lambda_a+\CQ_\beta^{-1}\Lambda_b \CQ_\beta$.\footnote{The subscripts $c$ and $f$ are there to indicate the classical and free convolutions, respectively. We refer to \cite{movassagh2010isotropic, PhysRevLett.107.097205} for a general discussion on these two extreme cases and the cases in between, when the matrices are not commutating, and also not asymptotically free - situation dubbed as the quantum convolution in \cite{movassagh2010isotropic}, keeping in mind the fact that realistic quantum operators show this behaviour.} 

Since the self-adjoint operator $\BC$ is assumed to have a compact DOS, the knowledge of the moments $m_k=\varphi(\BC^k)=\varphi\Big((\BA+\BB)^k\Big)$ is enough to determine $\rho_\BC$. The strategy for obtaining the DOS of $\BC$ when $\BA$ and $\BB$ are not free (or commutative) is to look into these individual moments ($m_n$) and compare the cases when $\BA$ and $\BB$ are free or commutative (classical) or when in-between (which we can call the `quantum' following \cite{movassagh2010isotropic}).  

Let us first compute the first three moments of $\BC$. To this end, we make the following generic \textit{assumptions}: 
\begin{enumerate}
    \item  The matrix $\BA$ is a diagonal matrix in the basis given (this is not a loss of generality). 
    \item The matrices $\BA$ and $\BB$ are independent (which implies that the elements of the diagonal matrices ($\Lambda_a$ and $\Lambda_b$) are random variables drawn independently from given distributions).
     \item  The matrix $\BB$ is such that the joint probability distribution of its eigenvalues and the joint distribution of the components of the eigenvectors are independent of each other (this is true, e.g., for the cases of classical  Gaussian random matrices \cite{Potters, livan2018introduction, Mehta}).
     \item $\CU$ is a random matrix whose elements have zero mean and variance $1/N$ (here $\CU$ stands for either of the three matrices $\Pi$ (for classical convolution), $\CQ_\beta$ (for free convolution) or $\CQ$ (for the general case).
\end{enumerate}

Now we have the following expressions  for the first three moments (recall that the $k$th moment is defined as $m_k=\avg{(\Lambda_a+\CU^{-1}\Lambda_b \CU)^k}$),  
\begin{align}
    m_1 &=\avg{\Lambda_a+\Lambda_b}~,\\
    m_2 &= \avg{\Lambda_a^2 +  2 \Lambda_a \CU^{-1} \Lambda_b \CU + \Lambda_b^2}~,\\
    m_3&=\avg{\Lambda_a^3 + 3 \Lambda_a^2 \CU^{-1}\Lambda_b \CU + 3 \Lambda_a \CU^{-1}\Lambda_b^2 \CU +\Lambda_b^3}~.
\end{align}
Our goal is to show that the first three moments are the same in all three cases, irrespective of $\CU$.
We can already see that the first moment $m_1$ does not depend on $\CU$. To prove this statement for the second and third-order moments, we consider the quantity $f_{k_1,k_2}=\avg{\Lambda_a^{k_1} \CU^{-1}\Lambda_b^{k_2} \CU}$, where $k_1>0, k_2>0$ are integers. By using the independence of the distribution of the eigenvalues and eigenvectors, and using the fact that the distribution of the elements of $\Lambda_a$ and $\Lambda_b$ are also independent, we can write $f_{k_1,k_2}= \avg{\Lambda_a^{k_1}}\avg{\Lambda_b^{k_2}}$, and hence, $f_{k_1,k_2}$ is independent of $\CU$. Since $m_2$ and $m_3$ depend on $f_{k_1,k_2}$ for various $k_1$ and $k_2$, the last statement shows that they also do not depend on $\CU$, thus are the same for free, classical and quantum cases. 

A similar expansion of the fourth moment readily shows that it contains a term of the form $(\Lambda_a\CQ^{-1}\Lambda_b \CQ)^2$ coming from the crossing partition in the moment expansion, which starts to appear in the fourth order \cite{nica2006lectures, speicher2019lecture}. Of course, in the asymptotic free case ($N \rightarrow \infty)$, the crossing partitions do not contribute to the moment expansion; however, in the general case of quantum convolution, the crossing partitions do contribute to the moments, making the fourth moments different in three cases.  Note that, as we have discussed in section \ref{sec_dos_anderson}, for the Anderson model with nearest neighbour interaction and on-site disorder, the difference between the free probability and quantum approximations starts to appear at the eighth moment; however, in general, the fourth moments are already distinct. 

\paragraph{Matching of the fourth moments.} 
The simplest way of quantifying the difference between the moments is to consider the difference between the fourth moments for the classical convolution and the other two cases, which can be rewritten as the difference between the crossing terms (here $\CU$ stands for $\CQ$ and $\CQ_\beta$),
\begin{equation}
    d_4= \avg{(\Lambda_a+\Pi^{-1}\Lambda_b \Pi)^4} - \avg{(\Lambda_a+\CU^{-1}\Lambda_b \CU)^4} = 2 \avg{(\Lambda_a\Pi^{-1}\Lambda_b \Pi)^2 - (\Lambda_a\CQ^{-1}\Lambda_b \CQ)^2}~.
\end{equation}

The procedure advocated in \cite{movassagh2010isotropic, PhysRevLett.107.097205,movassagh2017eigenvalue} makes the hypothesis that the fourth moment of $\BC$ can be obtained as a one-parameter linear combination of fourth moments of the classical and free cases, 
\begin{equation}\label{4moments}
 \avg{\BC^4} = p \avg{\BC_f^4} + (1-p) \avg{\BC_c^4} ~, ~~\text{i.e.,}~ m_4= p m_4^f + (1-p)m_4^c ~~\text{with}~~ 0 \leq p \leq 1~,
\end{equation}
which, in terms of distributions, states that the probability distribution 
$\rho_p(\lambda)$ has the same first four moments as the theoretical distribution $\rho_q(\lambda)$ where 
\begin{equation}
    \rho_p(\lambda)=p \rho_c (\lambda)+(1-p) \rho_f(\lambda) \approx \rho_q(\lambda)~.
\end{equation}
The upper and lower limits of $p$, correspond to two extreme cases,  respectively, when the operators $\BA$ and $\BB$ are free ($p=1$) or commutating ($p=0)$. 
The last expression can be utilised, along with the canonical representation of the asymptotic free variables, to obtain a general expression of the parameter $p$\footnote{We emphasise that this expression for $p$ is generic, and is true irrespective of the assumptions listed above.} \cite{movassagh2010isotropic, movassagh2017eigenvalue},
\begin{equation}\label{p_moments}
    p= \frac{\avg{\BC_c^4} -\avg{\BC^4}}{\avg{\BC_c^4} - \avg{\BC_f^4} }= \frac{\avg{\BA^2 \BB^2} - \avg{ (\BA\BB)^2}}{\avg{\BA^2 \BB^2}  - \avg{\big( \BA \CQ_\beta^{-1}\Lambda_b \CQ\big)^2}}~.
\end{equation}

We start the computation of $p$ by computing the second term in the denominator.  Denoting the eigenvalues of $\BA$ and $\BB$ as $a_i$ and $b_i$, respectively, we have,\footnote{For notational convenience, we have removed the subscript $\beta$ when writing the matrix elements, i.e., $\CQ_{ij}$ denotes a matrix element of $\CQ_\beta$. } 
\begin{equation}\label{4thfree}
\begin{split}
    \text{Tr}\Big[\big( \BA \CQ_\beta^{-1}\Lambda_b \CQ\big)^2\Big] = \sum_{i,j} a_i^2 b_j^2 |\CQ_{ij}|^4 + \sum_{i, j \neq n} a_i^2 b_jb_n |\CQ_{ij}|^2 |\CQ_{in}|^2 + \sum_{i\neq m,j} a_i a_m b_j 
    |\CQ_{ij}|^2 |\CQ_{mj}|^2 \\
    + \sum_{i\neq m,j \neq n} a_i a_m b_jb_n (\CQ^{-1})_{ij} \CQ_{jm}  (\CQ^{-1})_{mn} \CQ_{ni}~.
    \end{split}
\end{equation}

To proceed further, we \textit{assume} that the elements of the matrix $\BA$ (which we have already assumed to be diagonal) are drawn from independent distributions with zero mean and unit variance. 
Since the matrices $\BA$ and $\BB$ are independent, this implies that
after the ensemble average is performed in \eqref{4thfree}, the last two terms vanish. 

Furthermore, since, according to the third assumption listed above, the matrix $\BB$ is such that the joint probability distribution of its eigenvalues and the joint distribution of the components of the eigenvectors are independent of each other, we have
\begin{equation}\label{4thfree2}
   \mee \text{Tr}\Big[\big( \BA \CQ_\beta^{-1}\Lambda_b \CQ\big)^2\Big] = \sum_{i,j} \mee(b_j^2 ) \mee( |\CQ_{ij}|^4) + 
   \sum_{i, j \neq n} \mee(b_jb_n ) \mee(|\CQ_{ij}|^2 |\CQ_{in}|^2 )~. 
\end{equation}

Next we compute $\mee \text{Tr}[\BA^2 \BB^2] $. Using the first two assumptions, we have\footnote{Here $b_{ij}$ denotes the elements of the matrix $\BB$. $b_{ii}$ are the diagonal elements, while $b_i$ denotes the eigenvalues of this matrix, i.e., elements of $\Lambda_b$.  }  $\mee \text{Tr}[\BA^2 \BB^2] = \sum_{i,j} \mee(a_i^2) \mee (|b_{ij}|^2) =  \sum_{i,j} \mee (|b_{ij}|^2) $. Similarly, to compute $\mee \text{Tr}[(\BA \BB)^2] $ we notice that, since, $\mee (a_i a_j) =\delta_{ij} \mee (a_i^2)=\delta_{ij}$, one has 
$\mee \text{Tr}[(\BA \BB^2)^2] =\sum_i \mee |b_{ii}|^2 $. 

To make further progress towards the computation of the parameter $p$, we shall consider a specific form for the matrix $\BB$. Below, we illustrate a specific case where it is possible to obtain $p$ analytically. For further examples, both of analytical computation of $p$, and its comparison with numerical computations (including the case of realistic spin chains with local interactions), we refer to \cite{movassagh2010isotropic, PhysRevLett.107.097205, movassagh2017eigenvalue}. 

\paragraph{Example: $\BB$ is a Gaussian random matrix.}
We assume $a_i$s are drawn from independent Gaussian distributions with variance unity, and $\BB$ is an $N \times N$ random matrix drawn from the Gaussian orthogonal or the Gaussian unitary ensemble. In the standard construction of these random matrices by sampling elements from independent Gaussian distributions \cite{Potters, livan2018introduction}, we take the variance of the diagonal elements to be unity, and that of the off-diagonal elements to be $\beta/2$, $\beta$ indicating the Dyson index. Note that the assumptions listed above are satisfied. 

For this choice of $\BB$, we have, 
\begin{equation}
    \mee \text{Tr}[\BA^2 \BB^2]=  \sum_{i,j} \mee (|b_{ij}|^2) = N+N(N-1)\frac{\beta}{2}~,~~~\text{and}~~~\mee \text{Tr}[(\BA \BB)^2] = N~.
\end{equation}
Next, we need to  calculate $\mee(b_i^2)$ and $\mee(b_ib_j), i\neq j$ in  \eqref{4thfree2}.  First to obtain $\mee(b_i^2)$, we note that, since $\BB$ is self-adjoint, $\mee(b_i^2)$  can be related to the Frobenius norm through the relation (which is nothing but the second moment of the matrix $m_2=\varphi(\BB^2)$)
\begin{equation}
    N^{-1}\sum_i b_i^2= ||\BB||_F^2 = N^{-1}\sum_{ij} |b_{ij}|^2~,
\end{equation}
where the last relation is the standard definition of the Frobenius norm \cite{ Potters,amir1958singular}. Taking the ensemble average of both sides, we have $\mee(b_i^2) = \mee ||\BB||_F^2= 1+ (N-1)\beta/2  $. On the other hand,  to compute $\mee(b_ib_j)$ for $ i \neq j$, we note that 
$\text{Tr}[\BB]^2=\sum_i b_i^2+\sum_{i\neq j} b_i b_j$, and $\text{Tr} [\BB]$ is a random variable with mean zero and variance $N$. Hence, 
\begin{equation}
    \sum_{i\neq j} \mee (b_i b_j) = \mee \text{Tr}[\BB]^2 -  \sum_i \mee(b_i^2)  ~~~\text{so that}~~~ \mee (b_ib_j)=-\frac{\beta}{2}~.
\end{equation}
Finally, $\mee( |\CQ_{ij}|^4)$ and $\mee(|\CQ_{ij}|^2 |\CQ_{in}|^2 )$ in \eqref{4thfree2} can be obtained using the standard formulas of the Weingarten calculus \cite{collins2022weingarten}.

Substituting these results in \eqref{4thfree2} we have $\mee \text{Tr}\Big[\big( \BA \CQ_\beta^{-1}\Lambda_b \CQ\big)^2\Big] =N$, and hence $p=1$ from \eqref{p_moments}. The result $p=1$ is expected due to the fact that, since $\BB$ is a Gaussian random matrix, it is asymptotically free from the random diagonal matrix $\BA$, hence $m_4$ is the same as $m_4^f$ in \eqref{4moments}.  

A slightly more complicated example where $p$ can be obtained analytically, and is not unity, is the case where the matrix $\BB$ consists of $k$ different square blocks of size $L \times L$, with $kL=N$, and each block is an independent GOE/GUE random matrix. The computation of $p$ presented above can be extended to this case, and we refer to \cite{movassagh2017eigenvalue} for details.

\section{Free compression and decompression of a distribution}\label{free_compression}
The purpose of this section is to review the procedure of \textit{free compression} \cite{nica2006lectures} and its inverse, known as the \textit{free decompression}, and their usefulness in estimating the DOS of a large matrix. 
First, we go through the concept of free compression of a matrix, and its inverse, the free decompression, which was introduced in a recent work \cite{ameli2025spectral}, and then subsequently discuss how the Cauchy transform of the DOS of the matrix changes under the free (de)compression.

To begin with, consider an infinite sequence of matrices of increasing size, $\{\BA_1, \BA_2, \cdots\}$, where $\BA_n \in \mrr^{n \times n}$ and has a well-defined convergent DOS in the limit $n \rightarrow \infty$,\footnote{As before, this implies the moments $m_k=\varphi(\BA_n^k)$ are convergent in the limit $n \rightarrow \infty$ for any integer $k \geq1$.} so that a given fixed matrix $\BA$ which we are interested in can be thought of as an element of this sequence. 
It is convenient to construct this sequence such that, for each value of $n$, the $n$th matrix of the sequence $\BA_n$ appears as the top left submatrix of the next matrix of the sequence, i.e., $\BA_{n+1}$. Therefore, going from $n$th matrix of the sequence to the next one, a new row and a column appear, and furthermore, the construction implies that for fixed values of $i,j \leq n$, the $(i,j)$th element of the matrix $\BA_n$ is constant and the same as that of the matrix $\BA$, i.e., $(\BA_n)_{ij}=(\BA)_{ij}$.   

Now, given the $n$th matrix of the sequence alone, one can not, a priori, predict the new row and column the next matrix of the sequence would have. A way of connecting elements of the sequence, while keeping the eigenvalue spectrum invariant along the way, is to perform a similarity transformation of $\BA_n$ by a random matrix. The most convenient choice is to perform this random similarity transform by a Haar random permutation matrix $\BP$.  After a similarity transform by this matrix, the elements of $\BA_n$ are given by $\BA_n^\sigma = (\BA_{\sigma(i)\sigma(j)})_{i,j}^n$ where $\sigma$ denotes an element of the group of permutation on $\{1,2, \cdots n\}$.  For our present purposes, we note an important result, which states that the sequence of matrices obtained by this similarity transformation is asymptotically free in the limit $n \rightarrow \infty$ \cite{nica1993asymptotically}. 

Keeping the above construction in mind, we consider the following question. Given a matrix $\BA$ (which can be deterministic), how can one obtain the DOS of a randomly sampled submatrix of dimension, say $n\times n$, approximately, and can this process be performed in reverse?  These are addressed by the following procedures, known as the free compression and decompression.   

Let us denote the top-left $n_s\times n_s$ submatrix of $\BA_n^\sigma$ taken from the transformed sequence above to be $\BB_{n_s}$. If the DOS of the matrix $\BA_n$ converges to the distribution $\rho_{\BA_n}$, then the DOS of the matrix $\BB_{n_s}$ is said to be the \textit{free compression} of the distribution $\rho_{\BA_n}$,, which we denote here as  $\rho_{\BA_{n_s}}$.
There exists a quite simple but important relationship between the $R$-transforms of these two distributions \cite{nica1996multiplication}, namely,  \textit{the $R$-transform of the DOS of the submatrix constructed from a proportion $\alpha$ of (top-left) rows and columns is given by the $R$-transfom of the DOS original matrix with scaling the argument $z$ by $\alpha$, i.e., $R_{n_s}(z)=R_n(\frac{n_s}{n} z)$, where $\alpha=n_s/n$. }

It was observed in a recent work \cite{ameli2025spectral} that one can, in principle, perform the procedure of free compression in reverse, i.e., gain information about the DOS of the large matrix $\BA_n^\sigma$ of the sequence from a submatrix. When written in terms of the $R$-transforms of the DOS of the larger and smaller matrices,  the resulting procedure, called the \textit{free decompression} in \cite{ameli2025spectral}, results in the following relation, $R_{n}(z)=R_{n_s}(\frac{n}{n_s} z)$.  
Free decompression was used in this work to obtain the approximate DOS of different random matrix ensembles, and we refer the reader to this work for more details. Here we will be interested in understanding how the process of free compression and its inverse affect the Cauchy transform of the DOS, and explicitly computing the compressed distribution for some simple, but illustrative examples. 

\textbf{Notation.} To simplify the notation, from now on, we shall denote 
the distribution $\rho_{\BA_n}(\lambda)$ as $\rho_{n}(\lambda)\equiv\rho(\lambda)$, and $\rho_{\BA_{n_s}}(\lambda)$ as $\rho_{n_s}(\lambda)$ or $\rho_{\alpha}(\lambda)$, where $\alpha=n_s/n$, and similar such subscripts for other transformations of the DOS. This is to indicate that these transformations are to be calculated from the original and compressed distributions, without always explicitly referring to the operator (or the random variable belonging to some non-commutative probability space) it might be associated with. Furthermore, in this section $\alpha$ always denotes the free compression parameter, which should not be confused with the perturbation parameter $\alpha$ used in Section \ref{dos_perturb}.

\subsection{Variation of the Cauchy transform under free compression}
As we have seen above, the free compression of a matrix drawn from $\BA_n^\sigma$ corresponds to a simple rescaling of the argument of the $R$-transform. However, as we have mentioned in section \ref{additive_conv}, the $R$-transform is usually difficult to obtain analytically, since it involves inverting the Cauchy transform of the DOS. Therefore, for practical applications, it is quite beneficial to understand how the free compression and its inverse affect the Cauchy transform itself.   

To this end, we first note that the relation between the $R$-transform of a distribution and the inverse of its Cauchy transform (eq. \eqref{B_and_R1}) implies the following, 
\begin{equation}
    R_{\BA}(G_{\BA}(z))+\frac{1}{G_{\BA}(z)}=z~. 
\end{equation}
This, when applied to our present context (with the present notation), for a freely compressed matrix gives,
\begin{equation}
 R_\rho(\alpha G_{\rho_\alpha}(z))+  \frac{1}{G_{\rho_\alpha}(z)} =z~,
\end{equation}
where $G_{\rho_\alpha(z)}$ denotes the Cauchy transform of the asymptotic eigenvalue distribution $\rho_\alpha(\lambda)$ of the free compressed matrix. To eliminate $R_\rho(z)$, we rewrite it in terms of the corresponding $\CB_\rho(z)$, and the resulting expression, when acted on both sides with $G_\rho(z)$ gives the desired relation between the Cauchy transforms of the original and compressed distributions to be,
\begin{equation}\label{G_alpha}
    \alpha G_{\rho_\alpha}(z) = G_\rho\Big(z+\frac{1-\alpha}{\alpha}\frac{1}{G_{\rho_\alpha}(z)}\Big)~=\sum_{j}\frac{\theta^j}{j!~ (G_{\rho_\alpha}(z))^j}\partial_z^jG_\rho(z)~,
\end{equation}
where $\theta=(1-\alpha)/\alpha$. The second expression above can be obtained by performing a manipulation similar to the one leading to the final expression in \eqref{GC_exact}, whose $j=0$ term is the original (uncompressed) Cauchy transform.

\textbf{Example: Compressed Kesten-McKay distribution.}
As a simple example, consider the Kesten-McKay distribution given in \eqref{KM_dist}, of which the arcsine distribution is a special case ($\eta=2$).  It can be checked that the above relation for the Cauchy transform under free compression gives the following expression for the Cauchy transform of the compressed Kesten-McKay distribution
\begin{equation}\label{Cauchy_KM_com}
G^{KM}_{\rho_\alpha}(z) = \frac{z(\eta-2\alpha) -\eta\sqrt{z^2+4 \alpha(\alpha-\eta)}}{2\alpha(\eta^2-z^2)}~.
\end{equation}
It is easy to see that for $\alpha=1$ and $\eta=2$, we get back the Cauchy transform of the original arcsine distribution given in \eqref{Cau_tr_ss_arc}. The analytical expression for the compressed Kesten-McKay distribution obtained using the Stieltjes inversion formula \eqref{dos_from_G} is given by
\begin{equation}\label{comr_KM}
    \rho^{KM}_\alpha(\lambda)=\frac{\eta \sqrt{4\alpha(\eta-\alpha)-\lambda^2}}{2 \pi \alpha~(\eta^2-\lambda^2)}~,
\end{equation}
which has a support $|\lambda|\leq 2\sqrt{\alpha(\eta-\alpha)}$ over the real line.

It is also interesting to note that the final expression in \eqref{G_alpha} is similar to the series expansion for the Cauchy transforms in eqs. \eqref{GC_kappa}, \eqref{GC_kappa_2} for a free addition.\footnote{Note that, even though they are structurally somewhat `similar', the crucial difference is the presence of $G_{\rho_\alpha}(z)$ in the denominator of  \eqref{G_alpha}, so that, unlike the subordination formula for the free additive convolution of two different distributions, the 2nd term in the argument of $G_\rho$ in \eqref{G_alpha} can not arise from $R$-transform of any normalised probability distribution function (see also the discussion on the connection between the free addition and free compression later in this section).} Therefore, one can ask if it is possible to use  \eqref{G_alpha} in a perturbative manner to approximate the Cauchy transform under free compression for a generic DOS. To investigate this question, we first note that the parameter $\theta$ does not act as a perturbation parameter for any value of the compression parameter $\alpha$. In fact, $0<\theta<1$ only when  $1/2<\alpha<1$, hence only for this range of the compression parameter, one can think of consistently approximating the Cauchy transform $G_{\rho_\alpha}(z) $. However, due to the structure of this series expansion, where successive corrections to $G_{\rho_\alpha}(z)$ would be in the denominator, unlike the series considered in Section \ref{dos_perturb}, it is not straightforward to consistently compute all orders of corrections in $\theta$, even when $\theta<1$.  

\begin{figure}[h!]
		\centering
          \begin{tabular}{cc}
		\includegraphics[width=0.5\textwidth]{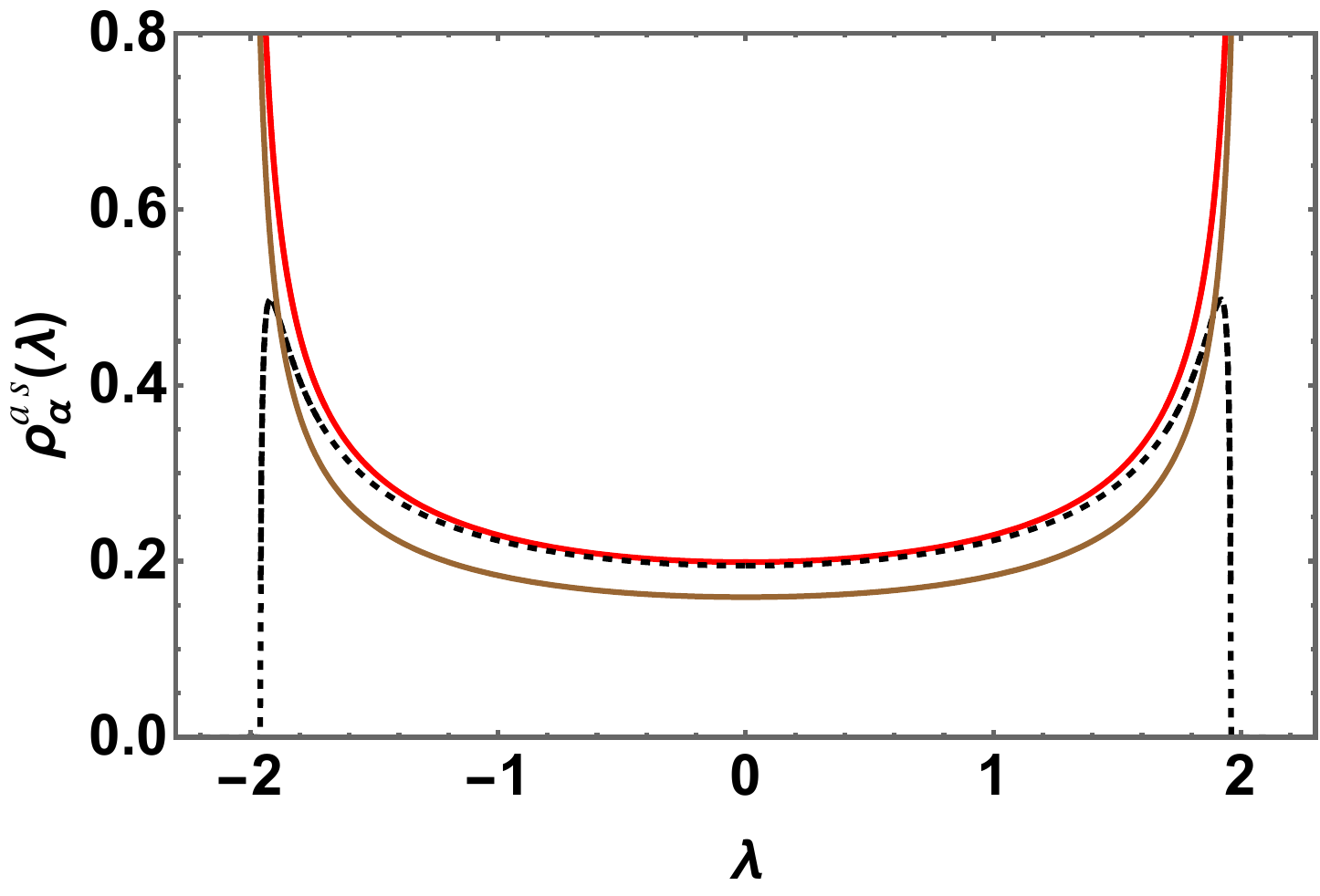}&
        \includegraphics[width=0.5\textwidth]{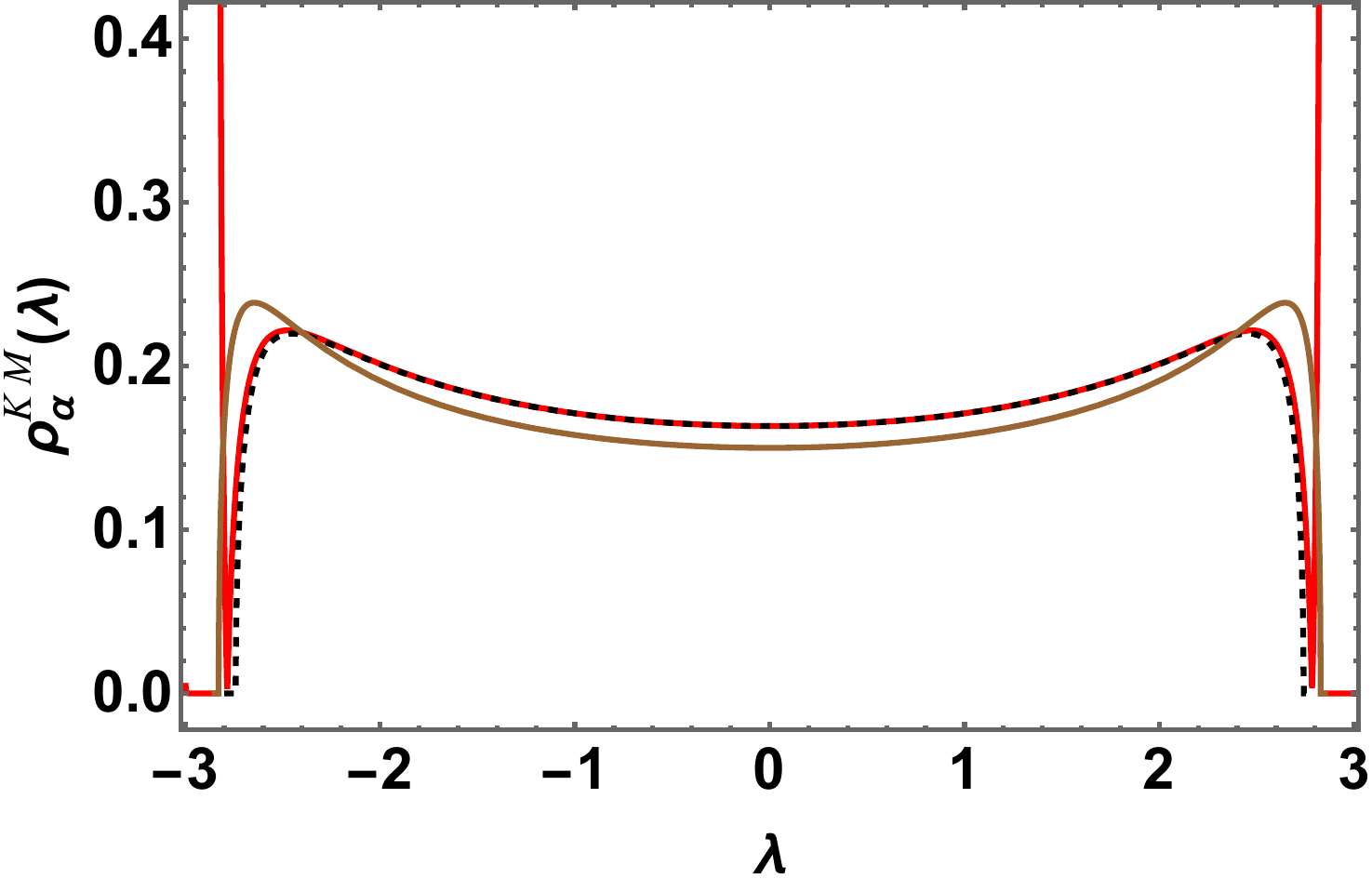}
        \end{tabular}
		\caption{Plots of the compressed arcsine and the Kesten-McKay distribution (with generic $\eta$), and their comparison with the DOS obtained from the approximate Cauchy transform in \eqref{Cauc_com_KM_app}. Plot in the left panel shows the free compression of the arcsine distribution with $\alpha=0.8$, while the plot in the right panel shows the free compression of the Kesten-McKay distribution with $\eta=3$, and $\alpha=0.89$. The red curves show the approximate DOS, while the black dashed curves represent the exact analytical compressed distributions \eqref{comr_KM}. The brown curve in the left panel indicates the uncompressed arcsine distribution, while the brown curve in the right panel shows the uncompressed Kesten-McKay distribution with $\eta=3$.}
		\label{fig:free_compression}
	\end{figure}

One can, nevertheless, unambiguously obtain the `first' correction term to the $G_n(z)$ under free compression, since $G_n(z)$ is independent of $\theta$. This is given by
\begin{equation}
    G_{\rho_\theta}(z) \approx (\theta+1) \Big(G_\rho(z)+ \theta \frac{\partial_zG_\rho(z)}{G_\rho(z)}\Big)~,~~\theta<1~.
\end{equation}
As an example, for the free compression of the Kesten-McKay distribution considered above, this can be evaluated to be
\begin{equation}\label{Cauc_com_KM_app}
    G^{KM}_{\rho_\theta}(z) \approx (\theta+1)\Bitd{\frac{z(\eta-2)-\eta \sqrt{4(1-\eta)+z^2}}{2(\eta^2-z^2)}+\theta\frac{z\sqrt{4(1-\eta)+z^2}-\eta(\eta-2)}{(\eta^2-z^2)\sqrt{4(1-\eta)+z^2}}}~.
\end{equation}
It can be easily checked that an expansion of $\alpha G^{KM}_{\rho_\alpha}(z)$ with $\alpha=1/(\theta+1)$ and $G^{KM}_{\rho_\alpha}(z)$ given in \eqref{Cauchy_KM_com} reproduces the expression in the square bracket above, and when $\theta=0$, we get back the Cauchy transform of the uncompressed Kesten-McKay distribution. To assess the accuracy of this approximation in predicting the compressed distribution, we consider two examples in Fig. \ref{fig:free_compression}: the compression of the arcsine distribution and the Kesten-McKay distribution with $\eta=3$. The approximate DOS (shown by the red curves in the two panels) provides a reasonably accurate description of the exact compressed distribution calculated from \eqref{Cauchy_KM_com} in the bulk of the distribution; however, the approximation becomes less accurate towards the edges. Moreover, as expected, as the value of the parameter $\alpha$ is decreased towards $1/2$ from unity (i.e., for a relatively higher amount of compression), the approximation gets worse. 

\paragraph{Partial differential equation satisfied by the Cauchy transform under free compression.}

Even though the formula in \eqref{G_alpha} relates the Cauchy transform of the DOS of the original matrix and the compressed one, we still need to solve complicated algebraic equations to obtain a solution for $G_{\rho_\alpha}(z)$ for a generic eigenvalue distribution. An alternative way of proceeding is to treat the parameter $\alpha$ also as a continuous variable (along with $z$) in the argument of the $R$ and Cauchy transforms, and try to understand how these two-variable functions $R(\alpha,z)$ or $G(\alpha,z)$ change as one varies $\alpha$ continuously. This is the approach taken in \cite{ameli2025spectral}, where the authors obtained a first-order partial differential equation (PDE) for $G(\alpha,z)$\footnote{In this reference, the authors have used the Stieltjes transform of the DOS,  which, as we have mentioned before, is just the negative of the Cauchy transform.}, which can then 
be solved by using the standard method of characteristics. Here, we derive this PDE for $G(\alpha,z)$, following a slightly different procedure.\footnote{We thank Kunal Pal for very useful discussions relevant to this section.}

First, we perform a change of variables, which will simplify the subsequent computations. We put $\alpha=e^u$, and consider the free (de)compression of the $R$-transform of a given distribution as a function of two variables, $u,z$, i.e., $R=R(u,z)$, similarly for the Cauchy transform and its inverse. For the free compression or decompression, we know that  $R(u,z)=R(ze^u)$, where $\alpha$ can represent any of the two processes, and $\alpha=1$, i.e., $u=0$ corresponds to the original matrix. Therefore, we do not add any additional subscripts to these functions.

We start by noting that, in terms of the $\CB(z)$-functions, the relation, 
$R(u,e^{-u}z)=R(0,z)$ can be rewritten as,
\begin{equation}
    \CB(u, e^{-u}z) - z^{-1}(e^u-1)=\CB(0,z)~.
\end{equation}
Now taking a partial derivative of both sides with respect to $u$, and denoting $e^{-u}z=y(u,z)$, for the moment, we have 
\begin{equation}
    \frac{\partial \CB(u, y(u,z))}{\partial u} +  \frac{\partial \CB(u, y(u,z))}{\partial y} \frac{\partial y(u,z)}{\partial u} - z^{-1}e^u=0~.
\end{equation}
Next, let us put $z \rightarrow z e^u$ in this equation, so that it reduces to a simple first-order PDE for the inverse of the Cauchy transform,
\begin{equation}
    \frac{\partial \CB(u, z)}{\partial u} -z  \frac{\partial \CB(u, z)}{\partial z} - \frac{1}{z} = 0~.
\end{equation}
To get an equation for the Cauchy transform itself, we employ the relation 
$G(u, \CB(u,z))=\CB(u, G(u,z))=z$. Differentiating this relation with respect to $z$ and $u$, respectively, to get $\partial_z\CB(u, z)$ and $\partial_u\CB(u, z)$, and subsequently substituting them in the above PDE, 
we have 
\begin{equation}
   \Big[ \frac{\partial G(u, \CB(u,z))}{\partial u} \Big]\Big[\frac{\partial \CB(u,z)}{\partial G(u, \CB(u,z))}\Big] + z \Big[\frac{\partial \CB(u,z)}{\partial G(u, \CB(u,z))}\Big] + z^{-1}=0~.
\end{equation}
Finally, evaluating this expression with $z \rightarrow G(u,z)$, we have the desired first-order PDE for the evolution of the Cauchy transform under free (de)compression,
\begin{equation}\label{Cauchy_pde}
    \frac{\partial G(u, z)}{\partial u} + G(u, z)+ \frac{1}{G(u, z)} \frac{\partial G(u, z)}{\partial z} = 0~.
\end{equation}

We will not discuss the solutions to this equation here, but rather refer to the original work of \cite{ameli2025spectral}, where solutions to this equation were obtained in terms of the characteristic curves, which can then be used to find an explicit solution. From this PDE, however, one simple question we can immediately answer is the following: Which distribution remains invariant under free compression, i.e., what is the solution for $G(z)$, when $\partial_uG(u,z)=0$. It is easy to check that, in this case, the solution to the above PDE is given by $G(z)=(z-c)^{-1}$, and the corresponding DOS $\rho(\lambda)=\delta(\lambda-c)$ does not have any continuous part, rather represents a Dirac mass at the value of the constant $\lambda=c$. This can, of course, be seen from the original relation for the $R$-transform for free compression, which in this case is just a constant $c$. 

\textbf{Examples.} We now discuss free compression of three distributions to illustrate this procedure, where the relevant Cuahcy transform of the compressed distribution can be found analytically. 

1. It can be checked that the Cauchy transform of a matrix with arcsine distribution as the eigenvalue distribution (e.g., a tridiagonal matrix with equal constant non-zero elements, as in the Anderson model discussed in section \ref{sec_dos_anderson}), which was derived in eq. \eqref{Cauchy_KM_com} (with $\eta=2$, for the case of the arcsine distribution) satisfies the above PDE \eqref{Cauchy_pde} (with $\alpha=e^u$). Also, the Cauchy transform in \eqref{Cauchy_KM_com} satisfies this PDE for other values of $\eta>2$ as well, since this represents a compressed Kesten-McKay distribution.

2. As the second example, consider the Bernoulli distribution in \eqref{Bernoulli_dist}, which can be thought of, e.g., as the eigenvalue distribution of a spin 1/2 Pauli operator acting on a particular site of a spin chain. Even though, as we have seen above, a single delta function distribution remains invariant under free compression, a distribution which is the sum of two such delta functions, representing two Dirac masses at different positions on the real line, does change under a free compression, since the corresponding $R$-transform is not a constant. The Cauchy transform of the Bernoulli distribution is given by $G(z)=z/(z^2-1)$, and  the $R$-transform can be evaluated to be given by \cite{speicher2019lecture}, $ R(z)=\frac{1}{2z} \big(-1+\sqrt{1+4 z^2}\big) $. The Cauchy transform of the Bernoulli distribution changes under a free compression as 
\begin{equation}\label{Cauchy_Ber_com}
G_{\rho_\alpha}(z) = \frac{z(2\alpha-1)+\sqrt{z^2+4 \alpha(\alpha-1)}}{2\alpha(z^2-1
)}~,
\end{equation}
and it can be verified that this satisfies the PDE in \eqref{Cauchy_pde} with $\alpha=e^u$.


3. As the final example, we consider the free compression of the measure induced by a sequence of polynomials $P(\lambda)$ that satisfy the following recursion relations with constant coefficients,
\begin{align}
    P_0(\lambda)&=c~,~P_1(\lambda)=\lambda-\gamma~,~~\\
    P_{n+1}(\lambda)&=(\lambda-a)P_{n}(\lambda)-bP_{n-1}(\lambda)~,~~\text{for}~~~n \geq 1~,
\end{align}
where $\gamma$ and $a$ are two real numbers, while the other two constants $b$ and $c$ are positive numbers. To keep the subsequent analytical expressions simple, henceforth, we shall set $c=1$, and $\gamma=0$, and assume that $a>0$. According to Favard theorem, up to a constant multiplication, there exists a compactly supported unique and positive measure on the real line, for which the above sequence of polynomials is orthogonal \cite{favard1935polynomes, chihara2011introduction, marcellan2001favard, golub2009matrices}. The expression for this probability measure induced by the sequence of orthogonal polynomials is given by \cite{cohen1984orthogonal, saitoh2001infinite}
\begin{equation}\label{dos_orthogonal}
    \rho^O(\lambda) = \frac{1}{2\pi} \frac{\sqrt{4 b -(\lambda-a)^2}}{b+a \lambda}~,
\end{equation}
which has a support over the range $|\lambda| \leq a \pm 2\sqrt{b}$ on the real line.\footnote{In general, the measure induced by the sequence of orthogonal polynomials with constant recurrence coefficients can have, apart from this continuous part, discrete Dirac masses on the real line \cite{cohen1984orthogonal, saitoh2001infinite}.} The Cauchy transform of this distribution, and the $R$-transform computed from it are given by
\begin{equation}
    G_{\rho^O}(\lambda)=\frac{a+z+\sqrt{(z-a)^2-4 b}}{2(az+b)}~,~~R_{\rho^O}(z)=\frac{b z }{1-a z}~.
\end{equation}
After a free compression, the Cauchy transform of the compressed distribution can be calculated to be 
\begin{equation}
    G_{\rho^O_\alpha}(\lambda)=\frac{a \alpha+z+\sqrt{(z-a \alpha)^2-4 b \alpha}}{2 \alpha(az+b)}~,
\end{equation}
which can be checked to satisfy the PDE \eqref{Cauchy_pde}. As is evident, the effect of the compression is to rescale the parameters $a$ and $b$ to $a \rightarrow a \alpha$ and $b \rightarrow b \alpha$. Hence, the expression of the compressed distribution is still given by the distribution in \eqref{dos_orthogonal} with these rescaled parameters.

\subsection{Free compression and free addition}

Free compression of the distribution of a random variable is closely related to the free additive convolution of the distribution of the variable with itself \cite{nica2006lectures}. 
Assume $x$ is a random variable (as before, we can take it to be an element of a $C^*$ probability space $(\CA, \varphi)$), and has a probability distribution $\rho$ having compact support on the real line. Consider a projection $P \in \CA$ which is free from $x$, and $\varphi(P)=\alpha$. Then a well-known result states that, the probability distribution of variables $PxP$ (which belongs to a compressed $C^*$ probability space)  is a distribution whose $n$th free cumulant is $1/\alpha$ times the $n$-th free cumulant of the scaled variable $\alpha x$ (see, e.g., \cite{nica2006lectures, speicher2019lecture} for the proof). This is the description of the compression by projection we are considering in this section in the non-commutative probability space, and the relation for the $ R$-transform for the free compression essentially follows from the last statement. The connection of this process of free compression and the free additive convolution can be drawn by remembering that scaling the free cumulants by a factor of $1/\alpha$ corresponds to $1/\alpha$-fold free additive convolution. 

To write this in terms of the $R$-transform, we recall that the scaling law for the $R$-transform says that\footnote{The notation $\alpha*\rho$ indicates the distribution of the scaled random variable, $\alpha x$. As mentioned previously, this is to associate these transformations directly with the probability distributions, without always referring to the underlying operator or the random variable in some non-commutative probability space (even though they exist by our assumption that these distributions are probability measures having compact supports on the real line).} $R_{\alpha*\rho}(z)=\alpha R_\rho(\alpha z)$. Now, combining this law with the relation $R_{\rho_\alpha}(z)=R_\rho(\alpha z)$ for a free compression of $\rho$, we see that the $R$-transforms of a scaling and a free compression are related through, $R_{\alpha*\rho}(z)=\alpha R_{\rho_\alpha}(z)$. On the other hand, since $R$-transform is additive under free additive convolution, considering $m$-fold free additive convolution of $\rho(\lambda)$ with itself, we have $R_{\rho^{\boxplus m}}(z)=m R_\rho(z)$. Making use of the last statements along with $m=1/\alpha$, we see that the $R$-transform of a free compression of $\rho(\lambda)$ by a factor of $\alpha$ is related to $R$-transform of the $m$ fold free convolution of the scaled distribution $\alpha*\rho$ by the following simple relation \cite{nica2006lectures, shlyakhtenko2020fractional}, 
\begin{equation}
    R_{\rho_\alpha}(z)= R_{(\alpha*\rho)^{\boxplus \frac{1}{\alpha}}}(z) =R_{\alpha*(\rho^{\boxplus \frac{1}{\alpha}})}(z)~.
\end{equation}
In terms of the random variable $x$, the last expression corresponds to a free addition of $m(=1/\alpha)$ scaled random variables $\alpha x$. 

Let us consider the case $\alpha=1/2$ separately, so that the RHS of the last relation corresponds to the usual (two-fold) free addition of $\rho(\lambda)$ with itself, along with a scaling transformation. It is straightforward to see that these two give rise to the relation \eqref{G_alpha} as follows. Subordination formulas for the free addition indicate that the Cauchy transform of $\rho^c(\lambda)=\rho^a(\lambda) \boxplus  \rho^a(\lambda)$ is 
\begin{equation}
    G_{\rho^c}(z) = G_{\rho^a}\Bitd{\frac{1}{2}\Bift{z+\frac{1}{G_{\rho^c}(z)}}}~.
\end{equation}
A further scaling of $\rho^c(\lambda)$ by a parameter $\alpha=1/2$ gives $G_{\alpha*\rho^c}(z) =2 G_{\rho^a}\Bitd{z+\frac{1}{G_{\alpha*\rho^c}(z)}}=G_{\rho^a_\alpha}(z)$, and so that the Cauchy transform reduces to the relation  \eqref{G_alpha} with $\alpha=1/2$. This procedure can be inductively continued for $\alpha=1/n$, where $n>1$ is a positive integer, to demonstrate that the relation between the Cauchy transforms of the original and the compressed distribution derived in \eqref{G_alpha} holds for these compression parameters.
As a simple example, if we take $\alpha=1/2$, the free compression of a Bernoulli distribution by this factor corresponds to two-fold free additive convolution of this distribution with itself, hence, up to a rescaling, gives the arcsine distribution. This can be directly checked from the formula for the Cauchy transform of the Bernoulli distribution in \eqref{Cauchy_Ber_com}, where the rescaling factor is $\alpha=1/2$, as expected. 

So far, in the above discussion on the connection between free compression and free additive convolution, we have been concerned with a single distribution and its scaling. As the next step, we can consider the interesting case of a free compression of a distribution, which is the free additive convolution of two different distributions.  As in Section \ref{dos_perturb}, let us assume that we perform a compression of $\rho^c(\lambda)=\rho^a(\lambda) \boxplus (\delta * \rho^b(\lambda))$,\footnote{To avoid possible confusion between the compression parameter $\alpha$ and the scaling/perturbation parameter of additive convolution of two different distributions, we have denoted the latter by $\delta$. } and derive two series expansions of the compressed distribution $\rho^c_\alpha(\lambda)$, combining the relation \eqref{G_alpha}, the subordination relations \eqref{subor_A}, \eqref{subor_B}, and the series expansion  \eqref{GC_kappa_2}. The first expansion can be derived straightforwardly by substituting \eqref{GC_kappa_2} in the last expression of \eqref{G_alpha} to have (recall that  $\theta=(1-\alpha)/\alpha$),
\begin{align}
    G_{\rho^c_\alpha}(z)&= (\theta+1) \sum_{j,k}\frac{(-1)^k\delta^k\theta^j}{j! k! (G_{\rho^c_\alpha}(z))^j} \partial^j_z \Bitd{\big[R_{\rho^b} \big(\delta~ G_{\rho^c}(z)\big)\big]^k~ \partial_z^k G_{\rho^a}(z)}~.
\end{align}
If we further assume that $\delta \ll 1$, then the above series can be used to get an approximate expression for the Cauchy transform of the compressed distribution at different orders of $\delta$. 

To derive a second expression, we utilise the subordination relation \eqref{subor_A} in the first expression of $G_{\rho^c_\alpha}(z)$ from \eqref{G_alpha} to get
\begin{equation}
     G_{\rho^c_\alpha}(z) = (\theta+1) G_{\rho^a}\Bift{z+\frac{\theta}{G_{\rho^c_\alpha}(z)}-\delta R_{\rho^b} \Big[\frac{\delta}{\theta+1}~ G_{\rho^c_\alpha}(z)\Big]}~.
\end{equation}
This expression is generic, and is a generalisation of \eqref{subor_A} and \eqref{G_alpha} for the free compression of a distribution which is a free additive convolution of two given distributions.
Now, following the steps leading to the relation \eqref{GC_exa_2}, we arrive at the series expansion,
\begin{equation}
     G_{\rho^c_\alpha}(z) = (\theta+1)  \sum_{k=0}^{\infty} \frac{[\partial_z^k G_{\rho^a}(z)]}{k!}\Bift{\frac{\theta}{G_{\rho^c_\alpha}(z)}+\delta R_{b} \Big[\frac{\delta}{\theta+1}~ G_{\rho^c_\alpha}(z)\Big]}^k~. 
\end{equation}
When the distribution $\rho_b(\lambda)$ is a Wigner semicircle, the above formula simplifies to the following
\begin{equation}
     G_{\rho^c_\alpha}(z) =  (\theta+1)    \sum_{k=0}^{\infty}\sum_{j=0}^{k} \frac{[\partial_z^k G_{\rho^a}(z)]}{j!(k-j)!} \theta^j\Bift{\frac{\delta^2 }{\theta+1}}^{k-j} ~ \bift{G_{\rho^c_\alpha}(z)}^{k-2j}~. 
\end{equation}

Even though it is beyond the scope of the present manuscript, it will be interesting to see whether the DOS of the quantum mechanical Hamiltonians with large dimension, which is the sum of two non-commutative operators (such as the Anderson model or the RP random matrix model), can be obtained by performing a free decompression after a free additive convolution of two known distributions, as we have discussed. We hope to report on this in a future work \cite{Inprep}.

\section{Discussions}\label{sec_discussion}

In this article, we have attempted to provide an overview of the computation of the DOS of quantum mechanical systems using the tools of free probability theory. Assuming that the Hamiltonian of the quantum system under consideration can be written as a sum of two non-commutating operators $H=\BA+\BB$, we can use the free convolution method to obtain an expression for the DOS of the system.  We have discussed examples of several quantum systems, including a class of local random matrix Hamiltonians with Haar-Ising interaction bonds, the RP random matrix Hamiltonian (and its generalisation with Gaussian $\beta$-ensemble), and different variants of the Anderson model, where one makes use of the $R$-transform and the subordination formulas for the Cauchy transform of the DOS to analytically compute the DOS of these models. Since the free additive convolution is a procedure that is hard to perform analytically apart from certain simple cases,  the perturbation scheme we have developed in Section \ref{dos_perturb} based on the subordination formulas is a very powerful method of obtaining, in a perturbative manner, different orders of corrections to the DOS of an initial operator when a second operator (with a scaling parameter), which is free from the initial operator, is added to it. We have illustrated the use of this formalism to calculate the DOS of the models mentioned above, and since the method is quite generic, we hope that it will find further applications in understanding the DOS of other complex interacting quantum systems beyond those we have considered here \cite{Inprep}. 

Besides these analytical methods, there exist several powerful numerical algorithms (particularly, based on the subordination formulas) which can be employed to approximate the DOS of different types of quantum mechanical operators to an excellent precision. Since the goal of the present manuscript is to focus solely on the analytical results, we have not discussed these numerical methods.  We refer the interested reader to \cite{olver, Camargo:2025zxr, Pollock:2025acf, ameli2025spectral, movassagh2017eigenvalue, chen2012partial, movassagh2010isotropic, rao2008polynomial} and references therein for further information and examples. 

Even though the underlying philosophy in applying the free probability has been that the component (non-commutating) operators of the Hamiltonian of a realistic quantum system are free, these being finite-dimensional matrices, are only `partially free' with respect to one another, i.e, their eigenvectors are not related by a completely random Haar unitary operator, and hence, the moments of the exact DOS match with the free probability prediction only up to a certain order. One way of quantifying the departure from free probability approximation is to compare the individual moments of the DOS, as we have described in Section \ref{sec_sum_AB}, and try to quantify the difference in a physically meaningful way. 

Furthermore, as we have seen in Section \ref{sec_dos_quantum}, given the Hamiltonian of a quantum system, the usefulness of the free probability approximation resides in the somewhat ad hoc procedure of choosing the component operators, $\BA$ and $\BB$ to be used in the free additive convolution process.  In fact, for a given quantum mechanical Hamiltonian, different choices of these operators can lead to quite different free probability predictions, and, apart from comparing the final result with the exact numerical DOS, there do not seem to be any universal guiding principles one can follow for this purpose.
In this context, we note that, as we have mentioned in the main text, for the perturbation scheme developed in section \ref{dos_perturb}, the interaction parameter with respect to which the perturbation scheme is developed provides a natural partition of the Hamiltonian into two component operators. In fact, depending on the strength of this parameter, one needs to suitably adjust the choices of the operators $\BA$ and $\BB$ for the scheme to work properly. 

In this manuscript, we have only discussed the properties of the eigenvalues
of the sum of two Hermitian operators, by assuming that they are large matrices and are free from one another, without considering the properties of the eigenvectors of $H=\BA+\BB$.  Specifically, as in Section \ref{dos_perturb},
if we consider the operator $\BB$ to be a perturbation to the operator $\BA$, then we can imagine that the eigenvectors of $H=\BA+\alpha \BB$ would be close to those of the operator $\BA$. Once again, free probability 
helps us to quantify this in a more precise manner. However, a detailed discussion of this is beyond the scope of this overview, and we refer to \cite{biane2003free} and the references therein for an introduction. 

Since its introduction in mathematics in the 1980s, the free probability theory has found an increasing number of applications in diverse branches of physics. In this paper, we have provided a brief review of a specific application of the powerful tools developed in this branch of mathematics in understanding the properties of the distribution of eigenvalues of complex, possibly many-body quantum systems with generic interactions. We hope that, from this manuscript, the reader will find an overview of the subject, where a large body of work has already been carried out by many authors, and it will motivate the reader to apply these mathematical tools to discover further applications of free probability in quantum physics. 

\begin{center}
	\bf{Acknowledgements}
\end{center}

We sincerely thank Hugo. A. Camargo, Yichao Fu, and, specifically, Viktor Jahnke, for many illuminating discussions on disordered quantum systems and free probability theory. 
We also thank Jaydeep Kumar Basak, Arpita Mitra, Debangshu Mukherjee, and Kunal Pal for many interesting questions and discussions; Klée Pollock and Jonathon Riddell for very useful correspondence and comments on a draft version of the manuscript.  We are grateful to Viktor Jahnke for carefully verifying the numerical results reported in Section \ref{dos_perturb}. 
KP would like to thank the hospitality of the Asia Pacific Centre for Theoretical Physics, Pohang, where part of this work was carried out. 

This work was supported by the Basic Science Research Program through the National Research Foundation of Korea (NRF) funded by the Ministry of Science, ICT \& Future Planning (NRF-2021R1A2C1006791), the framework of international cooperation program managed by the NRF of Korea (RS-2025-02307394), the Creation of the Quantum Information Science R\& D Ecosystem (Grant No. RS-2023-NR068116) through the National Research Foundation of Korea (NRF) funded by the Korean government (Ministry of Science and ICT), the Gwangju Institute of Science and Technology (GIST) research fund (Future leading Specialized Resarch Project, 2025) and the Al-based GIST Research Scientist Project grant funded by the GIST in 2025. This research was also supported by the Regional Innovation System \& Education(RISE) program through the (Gwangju RISE Centre), funded by the Ministry of Education(MOE) and the (Gwangju Metropolitan City), Republic of Korea (2025-RISE-05-001).

\bibliography{reference}{}

@article{Andersonlocal,
  title = {Absence of Diffusion in Certain Random Lattices},
  author = {Anderson, P. W.},
  journal = {Phys. Rev.},
  volume = {109},
  issue = {5},
  pages = {1492--1505},
  numpages = {0},
  year = {1958},
  month = {Mar},
  publisher = {American Physical Society},
  doi = {10.1103/PhysRev.109.1492},
  url = {https://link.aps.org/doi/10.1103/PhysRev.109.1492}
}

@article{Chenprl,
  title = {Error Analysis of Free Probability Approximations to the Density of States of Disordered Systems},
  author = {Chen, Jiahao and Hontz, Eric and Moix, Jeremy and Welborn, Matthew and Van Voorhis, Troy and Su\'arez, Alberto and Movassagh, Ramis and Edelman, Alan},
  journal = {Phys. Rev. Lett.},
  volume = {109},
  issue = {3},
  pages = {036403},
  numpages = {5},
  year = {2012},
  month = {Jul},
  publisher = {American Physical Society},
  doi = {10.1103/PhysRevLett.109.036403},
  url = {https://link.aps.org/doi/10.1103/PhysRevLett.109.036403}
}

@article{Welborn,
  title = {Densities of states for disordered systems from free probability},
  author = {Welborn, Matthew and Chen, Jiahao and Van Voorhis, Troy},
  journal = {Phys. Rev. B},
  volume = {88},
  issue = {20},
  pages = {205113},
  numpages = {9},
  year = {2013},
  month = {Nov},
  publisher = {American Physical Society},
  doi = {10.1103/PhysRevB.88.205113},
  url = {https://link.aps.org/doi/10.1103/PhysRevB.88.205113}
}

@article{chen2012partial,
  title={Partial freeness of random matrices},
  author={Chen, Jiahao and Van Voorhis, Troy and Edelman, Alan},
  journal={arXiv preprint arXiv:1204.2257},
  year={2012}
}

@article{movassagh2010isotropic,
  title={Isotropic entanglement},
  author={Movassagh, Ramis and Edelman, Alan},
  journal={arXiv preprint arXiv:1012.5039},
  year={2010}
}

@article{neu1993self,
  title={A self-consistent master equation and a new kind of cumulants},
  author={Neu, Peter and Speicher, Roland},
  journal={Zeitschrift f{\"u}r Physik B Condensed Matter},
  volume={92},
  number={3},
  pages={399--407},
  year={1993},
  publisher={Springer}
}

@article{BREZIN1996697,
title = "{Correlations of nearby levels induced by a random potential}",
    eprint = "cond-mat/9605046",
    archivePrefix = "arXiv",
    primaryClass = "cond-mat",
journal = {Nuclear Physics B},
volume = {479},
number = {3},
pages = {697-706},
year = {1996},
issn = {0550-3213},
doi = {https://doi.org/10.1016/0550-3213(96)00394-X},
url = {https://www.sciencedirect.com/science/article/pii/055032139600394X},
author = {E. Brézin and S. Hikami}
}

@article{zinnjustin1999adding,
  title={Adding and multiplying random matrices: a generalization of Voiculescu’s formulas},
  author={Zinn-Justin, Paul},
  journal={Physical Review E},
  volume={59},
  number={5},
  pages={4884},
  year={1999},
  publisher={APS}
}

@article{speicher1993free,
  title={Free convolution and the random sum of matrices},
  author={Speicher, Roland},
  journal={Publications of the Research Institute for Mathematical Sciences},
  volume={29},
  number={5},
  pages={731--744},
  year={1993},
  publisher={Research Institute forMathematical Sciences}
}

@article{gudowska2003free,
  title={Free random variables and molecular spectra},
  author={Gudowska-Nowak, Ewa and Kami{\'n}ska, Agnieszka and Papp, G{\'a}bor and Brickmann, J{\"u}rgen},
  journal={Physica A: Statistical Mechanics and its Applications},
  volume={325},
  number={1-2},
  pages={48--54},
  year={2003},
  publisher={Elsevier}
}

@article{gudowska1998bridged,
  title={Bridged-assisted electron transfer. Random matrix theory approach},
  author={Gudowska-Nowak, Ewa and Papp, G{\'a}bor and Brickmann, J{\"u}rgen},
  journal={Chemical physics},
  volume={232},
  number={3},
  pages={247--255},
  year={1998},
  publisher={Elsevier}
}

@article{ameli2025spectral,
  title={Spectral Estimation with Free Decompression},
  author={Ameli, Siavash and van der Heide, Chris and Hodgkinson, Liam and Mahoney, Michael W},
  journal={arXiv preprint arXiv:2506.11994},
  year={2025}
}

@article{voiculescu1991limit,
  title={Limit laws for random matrices and free products},
  author={Voiculescu, Dan},
  journal={Inventiones mathematicae},
  volume={104},
  number={1},
  pages={201--220},
  year={1991},
  publisher={Springer}
}

@article{Gopakumar:1994iq,
    author = "Gopakumar, Rajesh and Gross, David J.",
    title = "{Mastering the master field}",
    eprint = "hep-th/9411021",
    archivePrefix = "arXiv",
    reportNumber = "PUPT-1520",
    doi = "10.1016/0550-3213(95)00340-X",
    journal = "Nucl. Phys. B",
    volume = "451",
    pages = "379--415",
    year = "1995"
}

@article{Douglas:1994zu,
    author = "Douglas, Michael R.",
    title = "{Large N gauge theory: Expansions and transitions}",
    eprint = "hep-th/9409098",
    archivePrefix = "arXiv",
    reportNumber = "RU-94-72",
    doi = "10.1016/0920-5632(95)00431-8",
    journal = "Nucl. Phys. B Proc. Suppl.",
    volume = "41",
    pages = "66--91",
    year = "1995"
}

@article{Pappalardi:2022aaz,
    author = "Pappalardi, Silvia and Foini, Laura and Kurchan, Jorge",
    title = "{Eigenstate Thermalization Hypothesis and Free Probability}",
    eprint = "2204.11679",
    archivePrefix = "arXiv",
    primaryClass = "cond-mat.stat-mech",
    doi = "10.1103/PhysRevLett.129.170603",
    journal = "Phys. Rev. Lett.",
    volume = "129",
    number = "17",
    pages = "170603",
    year = "2022"
}

@article{PhysRevX.15.011031,
  title = {Designs via Free Probability},
  author = {Fava, Michele and Kurchan, Jorge and Pappalardi, Silvia},
  journal = {Phys. Rev. X},
  volume = {15},
  issue = {1},
  pages = {011031},
  numpages = {26},
  year = {2025},
  month = {Feb},
  publisher = {American Physical Society},
  doi = {10.1103/PhysRevX.15.011031},
  url = {https://link.aps.org/doi/10.1103/PhysRevX.15.011031}
}

@article{Vallini:2024bwp,
    author = "Vallini, Elisa and Pappalardi, Silvia",
    title = "{Long-time Freeness in the Kicked Top}",
    eprint = "2411.12050",
    archivePrefix = "arXiv",
    primaryClass = "cond-mat.stat-mech",
    month = "11",
    year = "2024"
}

@article{Pappalardi:2023nsj,
    author = "Pappalardi, Silvia and Fritzsch, Felix and Prosen, Toma\v{z}",
    title = "{Full Eigenstate Thermalization via Free Cumulants in Quantum Lattice Systems}",
    eprint = "2303.00713",
    archivePrefix = "arXiv",
    primaryClass = "cond-mat.stat-mech",
    month = "3",
    year = "2023"
}

@article{Fritzsch:2025arx,
    author = "Fritzsch, Felix and Claeys, Pieter W.",
    title = "{Free Probability in a Minimal Quantum Circuit Model}",
    eprint = "2506.11197",
    archivePrefix = "arXiv",
    primaryClass = "quant-ph",
    month = "6",
    year = "2025"
}

@article{Dowling:2025cxr,
    author = "Dowling, Neil and De Nardis, Jacopo and Heinrich, Markus and Turkeshi, Xhek and Pappalardi, Silvia",
    title = "{Free Independence and Unitary Design from Random Matrix Product Unitaries}",
    eprint = "2508.00051",
    archivePrefix = "arXiv",
    primaryClass = "quant-ph",
    month = "7",
    year = "2025"
}

@article{janik1997various,
  title={Various shades of Blue's functions},
  author={Janik, Romuald A and Nowak, Maciej A and Papp, Gabor and Zahed, Ismail},
  journal={arXiv preprint hep-th/9710103},
  year={1997}
}

@article{burda,
  title = {Multiplication law and $S$ transform for non-Hermitian random matrices},
  author = {Burda, Z. and Janik, R. A. and Nowak, M. A.},
  journal = {Phys. Rev. E},
  volume = {84},
  issue = {6},
  pages = {061125},
  numpages = {17},
  year = {2011},
  month = {Dec},
  publisher = {American Physical Society},
  doi = {10.1103/PhysRevE.84.061125},
  url = {https://link.aps.org/doi/10.1103/PhysRevE.84.061125}
}

@article{Engelhardt:1996da,
    author = "Engelhardt, M. and Levit, S.",
    title = "{Variational master field for large N interacting matrix models: Free random variables on trial}",
    eprint = "hep-th/9609216",
    archivePrefix = "arXiv",
    doi = "10.1016/S0550-3213(97)00043-6",
    journal = "Nucl. Phys. B",
    volume = "488",
    pages = "735--774",
    year = "1997"
}

@article{Wang:2022ots,
    author = "Wang, Jinzhao",
    title = "{Beyond islands: a free probabilistic approach}",
    eprint = "2209.10546",
    archivePrefix = "arXiv",
    primaryClass = "hep-th",
    doi = "10.1007/JHEP10(2023)040",
    journal = "JHEP",
    volume = "10",
    pages = "040",
    year = "2023"
}

@article{Vardhan:2025rky,
    author = "Vardhan, Shreya and Wang, Jinzhao",
    title = "{Free mutual information and higher-point OTOCs}",
    eprint = "2509.13406",
    archivePrefix = "arXiv",
    primaryClass = "quant-ph",
    month = "9",
    year = "2025"
}

@incollection{biane2003free,
  title={Free probability for probabilists},
  author={Biane, Philippe},
  booktitle={Quantum Probability Communications: QP--PQ (Volumes XI)},
  pages={55--71},
  year={2003},
  publisher={World Scientific}
}

@article{janiknonh,
  title = {Non-Hermitian random matrix models: Free random variable approach},
  author = {Janik, Romuald A. and Nowak, Maciej A. and Papp, G\'abor and Wambach, Jochen and Zahed, Ismail},
  journal = {Phys. Rev. E},
  volume = {55},
  issue = {4},
  pages = {4100--4106},
  numpages = {0},
  year = {1997},
  month = {Apr},
  publisher = {American Physical Society},
  doi = {10.1103/PhysRevE.55.4100},
  url = {https://link.aps.org/doi/10.1103/PhysRevE.55.4100}
}

@article{burdalevy,
  title = {Free random L\'evy matrices},
  author = {Burda, Zdzis\l{}aw and Janik, Romuald A. and Jurkiewicz, Jerzy and Nowak, Maciej A. and Papp, Gabor and Zahed, Ismail},
  journal = {Phys. Rev. E},
  volume = {65},
  issue = {2},
  pages = {021106},
  numpages = {5},
  year = {2002},
  month = {Jan},
  publisher = {American Physical Society},
  doi = {10.1103/PhysRevE.65.021106},
  url = {https://link.aps.org/doi/10.1103/PhysRevE.65.021106}
}

@article{Shtanko:2018oih,
    author = "Shtanko, Oles and Movassagh, Ramis",
    title = "{Stability of Periodically Driven Topological Phases against Disorder}",
    eprint = "1803.08519",
    archivePrefix = "arXiv",
    primaryClass = "cond-mat.str-el",
    doi = "10.1103/PhysRevLett.121.126803",
    journal = "Phys. Rev. Lett.",
    volume = "121",
    number = "12",
    pages = "126803",
    year = "2018"
}

@article{Venturelli:2022hka,
    author = "Venturelli, Davide and Cugliandolo, Leticia F. and Schehr, Gr\'egory and Tarzia, Marco",
    title = "{Replica approach to the generalized Rosenzweig-Porter model}",
    eprint = "2209.11732",
    archivePrefix = "arXiv",
    primaryClass = "cond-mat.dis-nn",
    doi = "10.21468/SciPostPhys.14.5.110",
    journal = "SciPost Phys.",
    volume = "14",
    number = "5",
    pages = "110",
    year = "2023"
}

@article{Pollock:2025acf,
    author = "Pollock, Kl{\'e}e and Kroth, Jonathan D. and Pagliaroli, Nathan and Iadecola, Thomas and Riddell, Jonathon",
    title = "{Energy dynamics in a class of local random matrix Hamiltonians}",
    eprint = "2502.05045",
    archivePrefix = "arXiv",
    primaryClass = "cond-mat.stat-mech",
    doi = "10.1103/5469-njfq",
    journal = "Phys. Rev. Res.",
    volume = "7",
    number = "3",
    pages = "033129",
    year = "2025"
}

@article{strang1999discrete,
  title={The discrete cosine transform},
  author={Strang, Gilbert},
  journal={SIAM review},
  volume={41},
  number={1},
  pages={135--147},
  year={1999},
  publisher={SIAM}
}

@article{collins2022weingarten,
  title={The weingarten calculus},
  author={Collins, Benoit and Matsumoto, Sho and Novak, Jonathan},
  journal={Notices of the American Mathematical Society},
  volume={69},
  number={05},
  pages={1},
  year={2022},
  publisher={American Mathematical Society (AMS)}
}

@article{Fava:2023pac,
    author = "Fava, Michele and Kurchan, Jorge and Pappalardi, Silvia",
    title = "{Designs via Free Probability}",
    eprint = "2308.06200",
    archivePrefix = "arXiv",
    primaryClass = "quant-ph",
    doi = "10.1103/PhysRevX.15.011031",
    journal = "Phys. Rev. X",
    volume = "15",
    number = "1",
    pages = "011031",
    year = "2025"
}

@article{Collins:2022klx,
    author = "Collins, Benoit and Yin, Zhi and Zhao, Liang and Zhong, Ping",
    title = "{The spectrum of local random Hamiltonians}",
    eprint = "2210.00855",
    archivePrefix = "arXiv",
    primaryClass = "math-ph",
    doi = "10.1088/1751-8121/acb4c8",
    journal = "J. Phys. A",
    volume = "56",
    number = "3",
    pages = "035201",
    year = "2023"
}

@article{charlesworth2021matrix,
  title={Matrix models for $\varepsilon$-free independence},
  author={Charlesworth, Ian and Collins, Beno{\^\i}t},
  journal={Archiv der Mathematik},
  volume={116},
  number={5},
  pages={585--600},
  year={2021},
  publisher={Springer}
}

@article{speicher2019lecture,
  title={Lecture Notes on" Free Probability Theory"},
  author={Speicher, Roland},
  journal={arXiv preprint arXiv:1908.08125},
  year={2019}
}

@article{voiculescu1986addition,
  title={Addition of certain non-commuting random variables},
  author={Voiculescu, Dan},
  journal={Journal of functional analysis},
  volume={66},
  number={3},
  pages={323--346},
  year={1986},
  publisher={Elsevier}
}

@article{olver,
  title={Numerical computation of convolutions in free probability theory},
  author={Olver, Sheehan and Rao Nadakuditi, Raj},
  journal={arXiv preprint  arXiv:1203.1958},
  year={2012},
}

@article{brezin1,
  title = {Correlation functions in disordered systems},
  author = {Br\'ezin, E. and Zee, A.},
  journal = {Phys. Rev. E},
  volume = {49},
  issue = {4},
  pages = {2588--2596},
  numpages = {0},
  year = {1994},
  month = {Apr},
  publisher = {American Physical Society},
  doi = {10.1103/PhysRevE.49.2588},
  url = {https://link.aps.org/doi/10.1103/PhysRevE.49.2588}
}

@article{brezin2,
  title = {Universal correlations for deterministic plus random Hamiltonians},
  author = {Br\'ezin, E. and Hikami, S. and Zee, A.},
  journal = {Phys. Rev. E},
  volume = {51},
  issue = {6},
  pages = {5442--5452},
  numpages = {0},
  year = {1995},
  month = {Jun},
  publisher = {American Physical Society},
  doi = {10.1103/PhysRevE.51.5442},
  url = {https://link.aps.org/doi/10.1103/PhysRevE.51.5442}
}

@article{PhysRevLett.107.097205,
  title = {Density of States of Quantum Spin Systems from Isotropic Entanglement},
  author = {Movassagh, Ramis and Edelman, Alan},
  journal = {Phys. Rev. Lett.},
  volume = {107},
  issue = {9},
  pages = {097205},
  numpages = {4},
  year = {2011},
  month = {Aug},
  publisher = {American Physical Society},
  doi = {10.1103/PhysRevLett.107.097205},
  url = {https://link.aps.org/doi/10.1103/PhysRevLett.107.097205}
}

@article{movassagh2017eigenvalue,
  title={Eigenvalue approximation of sums of Hermitian matrices from eigenvector localization/delocalization},
  author={Movassagh, Ramis and Edelman, Alan},
  journal={arXiv preprint arXiv:1710.09400},
  year={2017}
}

@article{amir1958singular,
  title={Singular values of a matrix},
  author={Amir-Mo{\'e}z, Ali R and Horn, Alfred},
  journal={The American Mathematical Monthly},
  volume={65},
  number={10},
  pages={742--748},
  year={1958},
  publisher={Taylor \& Francis}
}

@article{Jahnke:2025exd,
    author = "Jahnke, Viktor and Nandy, Pratik and Pal, Kuntal and Camargo, Hugo A. and Kim, Keun-Young",
    title = "{Free Probability approach to spectral and operator statistics in Rosenzweig-Porter random matrix ensembles}",
    eprint = "2506.04520",
    archivePrefix = "arXiv",
    primaryClass = "hep-th",
    reportNumber = "RIKEN-iTHEMS-Report-25",
    month = "6",
    year = "2025"
}

@article{neu1994spectra,
  title={Spectra of Hamiltonians with generalized single-site dynamical disorder},
  author={Neu, Peter and Speicher, Roland},
  journal={Zeitschrift f{\"u}r Physik B Condensed Matter},
  volume={95},
  number={1},
  pages={101--111},
  year={1994},
  publisher={Springer}
}

@article{neu1995rigorous,
  title={Rigorous mean-field model for coherent-potential approximation: Anderson model with free random variables},
  author={Neu, Peter and Speicher, Roland},
  journal={Journal of statistical physics},
  volume={80},
  number={5},
  pages={1279--1308},
  year={1995},
  publisher={Springer}
}

@article{neu1995random,
  title={Random matrix theory for CPA: generalization of Wegner's n-orbital model},
  author={Neu, Peter and Speicher, P},
  journal={Journal of Physics A: Mathematical and General},
  volume={28},
  number={3},
  pages={L79},
  year={1995},
  publisher={IOP Publishing}
}

@article{Wegner,
  title = {Disordered system with $n$ orbitals per site: $n=\ensuremath{\infty}$ limit},
  author = {Wegner, Franz J.},
  journal = {Phys. Rev. B},
  volume = {19},
  issue = {2},
  pages = {783--792},
  numpages = {0},
  year = {1979},
  month = {Jan},
  publisher = {American Physical Society},
  doi = {10.1103/PhysRevB.19.783},
  url = {https://link.aps.org/doi/10.1103/PhysRevB.19.783}
}

@article{Chen:2024free,
    author = "Chen, Hyaline Junhe and Kudler-Flam, Jonah",
    title = "{Free independence and the noncrossing partition lattice in dual-unitary quantum circuits}",
    eprint = "2409.17226",
    archivePrefix = "arXiv",
    primaryClass = "cond-mat.stat-mech",
    doi = "10.1103/PhysRevB.111.014311",
    journal = "Phys. Rev. B",
    volume = "111",
    number = "1",
    pages = "014311",
    year = "2025"
}

@article{AS,
    author = "Silva, Alessandro",
    title = "{Statistics of the Work Done on a Quantum Critical System by Quenching a Control Parameter}",
    doi = "10.1103/PhysRevLett.101.120603",
    journal = "Phys. Rev. Lett.",
    volume = "101",
    number = "12",
    pages = "120603",
    year = "2008"
}

@book{Mehta,
  title={Random Matrices},
  author={Mehta, M.L.},
  url={https://books.google.co.kr/books?id=vBHSAQAACAAJ},
  year={1991},
  publisher={Academic Press}
}

@article{dumitriu2002matrix,
    author = {Dumitriu, Ioana and Edelman, Alan},
    title = "{Matrix models for beta ensembles}",
    journal = {Journal of Mathematical Physics},
    volume = {43},
    number = {11},
    pages = {5830-5847},
    year = {2002},
    month = {11},
    issn = {0022-2488},
    doi = {10.1063/1.1507823},
    url = {https://doi.org/10.1063/1.1507823},
}

@book{forrester2010log,
  title={Log-gases and random matrices (LMS-34)},
  author={Forrester, Peter J},
  year={2010},
  publisher={Princeton university press}
}

@article{Hruza,
  title = {Coherent Fluctuations in Noisy Mesoscopic Systems, the Open Quantum SSEP, and Free Probability},
  author = {Hruza, Ludwig and Bernard, Denis},
  journal = {Phys. Rev. X},
  volume = {13},
  issue = {1},
  pages = {011045},
  numpages = {28},
  year = {2023},
  month = {Mar},
  publisher = {American Physical Society},
  doi = {10.1103/PhysRevX.13.011045},
  url = {https://link.aps.org/doi/10.1103/PhysRevX.13.011045}
}

@article{Wu:2023kin,
    author = "Wu, Shuang",
    title = "{Non-commutative probability insights into the double-scaling limit SYK model with constant perturbations: moments, cumulants and q-independence}",
    eprint = "2312.04297",
    archivePrefix = "arXiv",
    primaryClass = "math-ph",
    doi = "10.1088/1751-8121/ad65a6",
    journal = "J. Phys. A",
    volume = "57",
    number = "32",
    pages = "325203",
    year = "2024"
}

@article{pluma2022dynamical,
  title={A dynamical version of the SYK model and the q-Brownian motion},
  author={Pluma, Miguel and Speicher, Roland},
  journal={Random Matrices: Theory and Applications},
  volume={11},
  number={03},
  pages={2250031},
  year={2022},
  publisher={World Scientific}
}

@article{rao2008polynomial,
  title={The polynomial method for random matrices},
  author={Rao, N Raj and Edelman, Alan},
  journal={Foundations of Computational Mathematics},
  volume={8},
  number={6},
  pages={649--702},
  year={2008},
  publisher={Springer}
}

@article{capitaine2016spectrum,
  title={Spectrum of deformed random matrices and free probability},
  author={Capitaine, Mireille and Donati-Martin, Catherine},
  journal={arXiv preprint arXiv:1607.05560},
  year={2016}
}

@article{ostilli2012cayley,
  title={Cayley Trees and Bethe Lattices: A concise analysis for mathematicians and physicists},
  author={Ostilli, Massimo},
  journal={Physica A: Statistical Mechanics and its Applications},
  volume={391},
  number={12},
  pages={3417--3423},
  year={2012},
  publisher={Elsevier}
}

@article{Jindal:2024zcg,
    author = "Jindal, Siddharth and Hosur, Pavan",
    title = "{Generalized free cumulants for quantum chaotic systems}",
    eprint = "2401.13829",
    archivePrefix = "arXiv",
    primaryClass = "cond-mat.stat-mech",
    doi = "10.1007/JHEP09(2024)066",
    journal = "JHEP",
    volume = "09",
    pages = "066",
    year = "2024"
}

@article{Pollock:2025uem,
    author = "Pollock, Kl{\'e}e and Kroth, Jonathan D. and Riddell, Jonathon and Iadecola, Thomas",
    title = "{Group word dynamics from local random matrix Hamiltonians and beyond}",
    eprint = "2510.23716",
    archivePrefix = "arXiv",
    primaryClass = "cond-mat.stat-mech",
    month = "10",
    year = "2025"
}

@article{nica1993asymptotically,
  title={Asymptotically free families of random unitaries in symmetric groups.},
  author={Nica, Alexandru},
   journal = "Pacific Journal of Mathematics",
    volume = "157",
    number= "2",
    year="1993"
}

@article{nica1996multiplication,
  title={On the multiplication of free N-tuples of noncommutative random variables},
  author={Nica, Alexandru and Speicher, Roland},
  journal={American Journal of Mathematics},
  volume={118},
  number={4},
  pages={799--832},
  year={1996},
  publisher={Johns Hopkins University Press}
}

@article{Gill:2025upp,
    author = "Gill, Ankit and Kim, Keun-Young and Pal, Kunal and Pal, Kuntal",
    title = "{Geometry of quantum states in random matrix ensembles and the chaos-integrability transition}",
    eprint = "2507.13067",
    archivePrefix = "arXiv",
    primaryClass = "quant-ph",
    month = "7",
    year = "2025"
}

@article{kesten1959symmetric,
  title={Symmetric random walks on groups},
  author={Kesten, Harry},
  journal={Transactions of the American Mathematical Society},
  volume={92},
  number={2},
  pages={336--354},
  year={1959},
  publisher={JSTOR}
}

@article{mckay1981expected,
  title={The expected eigenvalue distribution of a large regular graph},
  author={McKay, Brendan D},
  journal={Linear Algebra and its applications},
  volume={40},
  pages={203--216},
  year={1981},
  publisher={Elsevier}
}

@article{livan2018introduction,
  title={Introduction to random matrices theory and practice},
  author={Livan, Giacomo and Novaes, Marcel and Vivo, Pierpaolo},
  journal={Monograph Award},
  volume={63},
  number={54},
  pages={914},
  year={2018}
}

@article{Fullgraf:2025tsw,
    author = {F{\"u}llgraf, Merlin and Gemmer, Jochen and Wang, Jiaozi},
    title = "{Scaling of free cumulants in closed system-bath setups}",
    eprint = "2511.11333",
    archivePrefix = "arXiv",
    primaryClass = "cond-mat.stat-mech",
    month = "11",
    year = "2025"
}

@article{cohen1984orthogonal,
  title={Orthogonal polynomials with a constant recursion formula and an application to harmonic analysis},
  author={Cohen, Joel M and Trenholme, Alice R},
  journal={Journal of functional analysis},
  volume={59},
  number={2},
  pages={175--184},
  year={1984},
  publisher={Academic Press}
}

@article{saitoh2001infinite,
  title={The infinite divisibility and orthogonal polynomials with a constant
recursion formula in free probability theory},
  author={Saitoh, Naoko and Yosnida, MWOAKI},
  journal={Probability and mathematical statistics },
  volume={59},
  pages={159–-170},
  year={2001}
}

@book{voiculescu1992free,
  title={Free random variables},
  author={Voiculescu, Dan V and Dykema, Ken J and Nica, Alexandru},
  volume={1},
  year={1992},
  publisher={American Mathematical Soc.}
}

@book{mingo2017free,
  title={Free probability and random matrices},
  author={Mingo, James A and Speicher, Roland},
  volume={35},
  year={2017},
  publisher={Springer}
}

@article{voiculescu1987multiplication,
  title={Multiplication of certain non-commuting random variables},
  author={Voiculescu, Dan},
  journal={Journal of Operator Theory},
  pages={223--235},
  year={1987},
  publisher={JSTOR}
}

@article{nica1998commutators,
  title={Commutators of free random variables},
  author={Nica, Alexandru and Speicher, Roland},
  eprint = "funct-an/9612001",
  year={1996}
}

@article{favard1935polynomes,
  title={Sur les polynomes de Tchebicheff},
  author={Favard, Jean},
  journal={CR Acad. Sci. Paris},
  volume={200},
  number={2052-2055},
  pages={11},
  year={1935}
}

@book{chihara2011introduction,
  title={An introduction to orthogonal polynomials},
  author={Chihara, Theodore S},
  year={2011},
  publisher={Courier Corporation}
}

@article{marcellan2001favard,
  title={On the “Favard theorem” and its extensions},
  author={Marcell{\'a}n, Francisco and {\'A}lvarez-Nodarse, Renato},
  journal={Journal of computational and applied mathematics},
  volume={127},
  number={1-2},
  pages={231--254},
  year={2001},
  publisher={Elsevier}
}

@book{golub2009matrices,
  title={Matrices, moments and quadrature with applications},
  author={Golub, Gene H and Meurant, G{\'e}rard},
  year={2009},
  publisher={Princeton University Press}
}

@article{shlyakhtenko2020fractional,
  title={Fractional free convolution powers},
  author={Shlyakhtenko, Dimitri and Jekel, Terence Tao and others},
  eprint = "2009.01882",
  archivePrefix = "arXiv",
  year={2020}
}

@article{collins2016random,
  title={Random matrix techniques in quantum information theory},
  author={Collins, Beno{\^\i}t and Nechita, Ion},
  journal={Journal of Mathematical Physics},
  volume={57},
  number={1},
  year={2016},
  publisher={AIP Publishing},
  eprint="1509.04689",
  archivePrefix = "arXiv"
}

@article{cortinovis2025computing,
  title={Computing free convolutions via contour integrals},
  author={Cortinovis, Alice and Ying, Lexing},
  journal={Random Matrices: Theory and Applications},
  volume={14},
  number={01},
  pages={2450024},
  year={2025},
  publisher={World Scientific},
  eprint="2305.01819",
  archivePrefix = "arXiv"
}

@article{Inprep,
  title={Density of states of quantum systems from free decompression},
  author={Kim, K.Y. and Pal, Kuntal.},
  journal={In Prepartion}
}

@book{Potters,
  title={A First Course in Random Matrix Theory},
  author={Potters, Marc and Bouchaud,  Jean-Philippe},
  year={2020},
  publisher={Cambridge University Press}
}

@article{rosenzweig1960repulsion,
  title={" Repulsion of Energy Levels" in Complex Atomic Spectra},
  author={Rosenzweig, Norbert and Porter, Charles E},
  journal={Physical Review},
  volume={120},
  number={5},
  pages={1698},
  year={1960},
  publisher={APS}
}

@article{kravtsov2015random,
  title={A random matrix model with localization and ergodic transitions},
  author={Kravtsov, VE and Khaymovich, IM and Cuevas, E and Amini, M},
  journal={New Journal of Physics},
  volume={17},
  number={12},
  pages={122002},
  year={2015},
  publisher={IOP Publishing}
}

@book{nica2006lectures,
  title={Lectures on the combinatorics of free probability},
  author={Nica, Alexandru and Speicher, Roland},
  volume={13},
  year={2006},
  publisher={Cambridge University Press}
}

@article{Camargo:2025zxr,
       author = {{Camargo}, Hugo A. and {Fu}, Yichao and {Jahnke}, Viktor and {Pal}, Kuntal and {Kim}, Keun-Young},
        title = "{Quantum Signatures of Chaos from Free Probability}",
      journal = {arXiv e-prints},
         year = 2025,
        month = mar,
          eid = {arXiv:2503.20338},
        pages = {arXiv:2503.20338},
          doi = {10.48550/arXiv.2503.20338},
archivePrefix = {arXiv},
       eprint = {2503.20338},
 primaryClass = {hep-th},
       adsurl = {https://ui.adsabs.harvard.edu/abs/2025arXiv250320338C}
}

@article{zee1996law,
  title={Law of addition in random matrix theory},
  author={Zee, Anthony},
  journal={Nuclear Physics B},
  volume={474},
  number={3},
  pages={726--744},
  year={1996},
  publisher={Elsevier}
}
\bibliographystyle{utphys}

\end{document}